\documentclass[12pt]{article}
\usepackage{putex}
\usepackage{graphicx}
\usepackage{epstopdf}
\usepackage{enumerate}
\usepackage{cite}

\newcommand{\abs}[1]{\left\lvert #1 \right\rvert}

\newcommand {\be} {\begin {equation}}
\newcommand {\ee} {\end {equation}}

\newcommand {\bes} {\begin {equation*}}
\newcommand {\ees} {\end {equation*}}

\newcommand{\es}[2] {\begin{equation} \label{#1} \begin{split} #2 \end{split} \end{equation}}

\newcommand{\Z}{\mathbb{Z}}
\newcommand{\R}{\mathbb{R}}
\newcommand{\C}{\mathbb{C}}

\DeclareMathOperator {\sh} {sinh}
\DeclareMathOperator {\ch} {cosh}

\begin{document}

\preprint{PUPT-2359}

\institution{PU}{Joseph Henry Laboratories$^1$ and Center for Theoretical Science,$^2$\cr
~~~~~~~~~~~~~~Princeton University, Princeton, NJ 08544}

\title{Multi-Matrix Models and Tri-Sasaki Einstein Spaces}

\authors{Christopher P.~Herzog,$^1$ Igor R.~Klebanov,$^{1, 2}$ Silviu S.~Pufu,$^1$\\[10pt] and Tiberiu Tesileanu$^1$}

\abstract{
Localization methods reduce the path integrals in ${\cal N}\geq 2$ supersymmetric Chern-Simons gauge theories on $S^3$ to multi-matrix integrals.  A recent evaluation of such a two-matrix integral for the ${\cal N}=6$ superconformal $U(N)\times U(N)$ ABJM theory produced detailed agreement with the AdS/CFT correspondence, explaining, in particular, the $N^{3/2}$ scaling of the free energy. We study a class of $p$-matrix integrals describing ${\cal N}=3$ superconformal $U(N)^p$ Chern-Simons gauge theories. We present a simple method that allows us to evaluate the eigenvalue densities and the free energies in the large $N$ limit keeping the Chern-Simons levels $k_i$ fixed. The dual M-theory backgrounds are $AdS_4\times Y$, where $Y$ are seven-dimensional tri-Sasaki Einstein spaces specified by the $k_i$. The gravitational free energy scales inversely with the square root of the volume of $Y$.   We find a general formula for the $p$-matrix free energies that agrees with the available results for volumes of the tri-Sasaki Einstein spaces $Y$, thus providing a thorough test of the corresponding $AdS_4$/CFT$_3$ dualities.  This formula is consistent with the Seiberg duality conjectured for Chern-Simons gauge theories.
}

\date{March 2011}


\maketitle


\section{Introduction and Summary}

The AdS/CFT correspondence \cite{Maldacena:1997re,Gubser:1998bc,Witten:1998qj} provides many predictions about the dynamics of strongly interacting field theories in various numbers of dimensions.
For the case of three dimensions, ref.~\cite{Klebanov:1996un} predicted that the number of low-energy degrees of freedom on $N$ coincident M2-branes scales as $N^{3/2}$ for large $N$.  Remarkably, this surprising prediction was recently confirmed~\cite{Drukker:2010nc} in the context of the ABJM construction \cite{Aharony:2008ug} of the $U(N)_k\times U(N)_{-k}$ Chern-Simons gauge theory on coincident M2-branes.  The paper \cite{Drukker:2010nc} was in turn based on~\cite{Kapustin:2009kz} where the methods of localization \cite{Pestun:2007rz} were shown to reduce the path integral of the Euclidean ABJM theory on $S^3$ to a matrix integral. This matrix model was solved in the large $N$ limit with $N/k$ kept fixed \cite{Marino:2009jd,Drukker:2010nc}, leading to precise tests of the AdS/CFT correspondence for Wilson loops and for the free energy (by which we mean minus the logarithm of the Euclidean partition function).
The exact solution of this matrix model is related by analytic continuation to a solution \cite{Halmagyi:2003ze} of another matrix model describing topological Chern-Simons theory on $S^3/\Z_2$; in particular, the formula for the resolvent has the same structure in the two cases. A generalization of the matrix model to the case where the Chern-Simons levels do not add up to zero was considered in \cite{Suyama:2010hr}.

The aim of our paper is to build on the major progress recently achieved in \cite{Kapustin:2009kz, Drukker:2010nc, Marino:2009jd} in several ways. In section 2 we revisit the matrix integral for the ABJM theory on $S^3$ and uncover the details of the eigenvalue distribution.
The matrix eigenvalues are located along the branch cuts of the resolvent used in \cite{Drukker:2010nc} and derived in \cite{Halmagyi:2003ze} for the $S^3/\Z_2$ model.
While the endpoints of the cuts can be read off directly from the resolvent, the cuts themselves are not simply parallel to the real axis, in contrast with the matrix model of \cite{Halmagyi:2003ze}.  In order to gain intuition for the location of the eigenvalues, we develop a numerical method for finite $N$ and $k$. This method allows us to access values of $N$ and $k$ that are large enough for the result to be a good approximation to the limit studied in \cite{Marino:2009jd,Drukker:2010nc}. Furthermore, we focus on the limit where $N$ is sent to infinity at fixed $k$ where the ABJM model is expected to be dual to the $AdS_4\times S^7/\Z_k$ background of M-theory. In this strong coupling limit, which is not of the 't Hooft type, we find analytically that the structure of the solution simplifies considerably.  An ansatz where the real parts of the eigenvalues scale with $\sqrt{N}$ allows us to calculate the free energy analytically.  Unlike in \cite{Drukker:2010nc}, our method does not rely on resolvents or mirror symmetry.  We confirm that the free energy scales as $N^{3/2}$ with the coefficient found in \cite{Drukker:2010nc}.

In section 3 we develop our analytic approach further and apply it to the large $N$ limit of matrix models describing quiver Chern-Simons gauge theories on $S^3$. We study explicitly a class of ${\cal N}=3$ superconformal $U(N)^p$ gauge theories with bifundamental matter, quartic superpotentials, and Chern-Simons levels $k_1, k_2, \ldots, k_p$ that sum to zero. These models were introduced in \cite{Imamura:2008nn} where their type IIB brane constructions were presented. The type IIB brane constructions involve $N$ D3-branes that are wrapped around a circle and intersect the $(1, q_1),(1, q_2), \ldots, (1, q_p)$ 5-branes located sequentially along the circle. The dual $AdS_4 \times Y$ M-theory backgrounds for these models, which involve certain seven-dimensional tri-Sasaki Einstein spaces $Y$, were conjectured in \cite{Jafferis:2008qz}.   The tri-Sasaki Einstein spaces $Y$ are defined to be bases of hyper-K\"ahler cones \cite{Boyer:1993, Boyer:1994,Boyer:1998sf}, and we take the Einstein metric on them to be normalized so that $R_{mn} = 6 g_{mn}$.  The $p$-matrix models for the gauge theories dual to $AdS_4 \times Y$ may be read off from \cite{Kapustin:2009kz}. In the large $N$ limit we calculate the eigenvalue densities for these matrix models and show that they are piecewise linear. This remarkably simple conclusion allows us to evaluate the coefficient of the $N^{3/2}$ scaling of the free energy as a function of the levels $k_i$ and compare it with the calculation on the gravity side of the AdS/CFT correspondence \cite{Emparan:1999pm,Drukker:2010nc}.
 For an arbitrary compact space $Y$ we find that the gravitational free energy is
  \es{FPreview}{
  F =  N^{3/2} \sqrt{\frac{2 \pi^6} {27\Vol(Y)}} \,.
  }
For $p=3$ the tri-Sasaki Einstein spaces $Y$ are the Eschenburg spaces \cite{Eschenburg} whose volumes were determined explicitly in \cite{Lee:2006ys}.  Our $3$-matrix model free energy is in perfect agreement with this volume formula.

Furthermore, we carry out calculations of the $p$-matrix model free energy and use them to conjecture an explicit general formula for the volumes via the AdS/CFT correspondence.  For a general $p$-node quiver with CS levels $k_a = q_{a+1} - q_a$, with $1 \leq a \leq p$ and $q_{p+1} = q_1$, we conjecture in section 4 that
 \es{ConjectureVolumePreview}{
  \frac{\Vol(Y)}{\Vol(S^7)} =
 \frac{ \sum_{(V,E) \in {\cal T}}  \prod_{(a,b)\in E} |q_a - q_b| }
  {\prod_{a = 1}^p \Bigl[ \sum_{b = 1}^p \abs{q_a - q_b} \Bigr] } \,,
 }
 where the sum in the numerator is over the set $\mathcal T$ of all trees
 (acyclic connected graphs) with $p$ nodes.  Such a tree $(V, E)$ consists of the vertices
 $V = \{1, 2, \ldots, p \}$ and $\abs E = p-1$ edges.   The volumes of the corresponding tri-Sasaki Einstein spaces $Y$ had previously been studied by Yee, who expressed them through an integral formula (eq.~(3.49) of \cite{Yee:2006ba}).  In the cases we have checked, our formula \eqref{ConjectureVolumePreview} is consistent with that of \cite{Yee:2006ba}.  Equation~\eqref{ConjectureVolumePreview} is invariant under permutations of the $q_a$, supporting the conjectured Seiberg duality for Chern-Simons theories with at least ${\cal N}=2$ supersymmetry \cite{Aharony:2008gk,Giveon:2008zn, Amariti:2009rb}, which may be motivated by interchanging different types of 5-branes in the type IIB brane constructions of these models.

Recent work \cite{Kapustin:2009kz,Drukker:2010nc,Jafferis:2010un} and the present paper hint at a special role that may be played by the Euclidean path integral of a three-dimensional conformal field theory on $S^3$.   This quantity may be analogous to the conformal anomaly coefficients in even dimensional CFTs.  Recall that the anomaly coefficients are very useful measures of the number of degrees of freedom.  For even dimensional theories with weakly curved dual backgrounds, these coefficients can be calculated using dual gravity in AdS space \cite{Henningson:1998gx} leading to precise tests of the AdS/CFT correspondence.  Such a definition of the number of degrees of freedom is not available for three-dimensional CFTs.  As mentioned already, the path integral on $S^3$ can be reduced to matrix integrals using supersymmetric localization methods \cite{Kapustin:2009kz}. Earlier work on gravity in Euclidean $AdS_4$ \cite{Emparan:1999pm} has pointed to the usefulness of the corresponding quantity: After adding certain surface counter-terms, the action becomes finite and appears to be unambiguous. The successful matching of this finite gravitational action with the path integral on $S^3$ in \cite{Drukker:2010nc} for the ${\cal N}=6$ ABJM theory and in the present paper for a class of ${\cal N}=3$ superconformal theories provides evidence for the usefulness of this quantity as a measure of the number of degrees of freedom.

 One may hope that the free energy on $S^3$ is also a useful quantity for nonsupersymmetric fixed points.\footnote{Of course, another nonsupersymmetric measure of the number of degrees of freedom, which is very useful physically, is the thermal free energy.}  For example, one could aim to match the large $N$ free energy on $S^3$ for the nonsupersymmetric example of AdS/CFT correspondence conjectured for the $O(N)$ sigma model in three dimensions \cite{Klebanov:2002ja}.

\section{ABJM Matrix Model}
\label{ABJM}

\subsection{Matrix Model Setup}

As shown in \cite{Kapustin:2009kz}, the partition function for ABJM theory on $S^3$ localizes on configurations where the auxiliary scalars $\sigma$ and $\tilde \sigma$ in the two ${\cal N} = 2$ vector multiplets are constant $N\times N$ Hermitian matrices.  Denoting the eigenvalues of $\sigma$ and $\tilde \sigma$ by $\lambda_i$ and $\tilde \lambda_i$, with $1 \leq i \leq N$, one can write the partition function as
 \es{ZABJM}{
  Z = {1 \over (N!)^2} \int \left(\prod_{i=1}^N \frac{d\lambda_i\, d\tilde \lambda_i}{(2 \pi)^2}\right)
   \frac{\prod_{i< j} \left(2 \sh \frac{\lambda_i - \lambda_j}{2} \right)^2 \left(2 \sh \frac{\tilde \lambda_i - \tilde \lambda_j}{2}\right)^2}
    {  \prod_{i, j} \left(2 \ch \frac{\lambda_i - \tilde \lambda_j}{2} \right)^2 }
    \exp \left( \frac{ik}{4 \pi} \sum_i (\lambda_i^2 - \tilde \lambda_i^2)  \right)\,,
 }
where $k$ is the Chern-Simons level, and the precise normalization was chosen as in \cite{Drukker:2010nc}.  The integration contour should be taken to be the real axis in each integration variable.  When the number $N$ of eigenvalues is large, the integral in eq.~\eqref{ZABJM} can be approximated in the saddle-point limit by $Z=e^{-F}$, where the ``free energy'' $F$ is an extremum of
 \es{FreeEnergy}{
  F(\lambda_i, \tilde \lambda_i) = -i {k \over 4 \pi} &\sum_j (\lambda_j^2 - \tilde \lambda_j^2)
  - \sum_{i < j} \log \left[  \left( 2 \sh {\lambda_i - \lambda_j \over 2} \right)^2
  \left( 2 \sh {\tilde \lambda_i - \tilde \lambda_j \over 2} \right)^2 \right] \\
    & \qquad {}+ 2 \sum_{i, j} \log \left(2 \ch {\lambda_i - \tilde \lambda_j \over 2} \right) + 2 \log N! + 2 N \log (2 \pi)
 }
with respect to $\lambda_i$ and $\tilde \lambda_i$.  The goal of this section is to compute the leading contribution to $F$ in such a large $N$ expansion while holding $k$ fixed.

Varying \eqref{FreeEnergy} with respect to $\lambda_j$ and $\tilde \lambda_j$ we obtain the saddle-point equations:
 \es{ABJMSaddle}{
- \frac{\partial F}{\partial \lambda_i} =   {i k\over 2\pi} \lambda_i - \sum_{j \neq i} \coth{\lambda_j - \lambda_i \over 2} + \sum_j \tanh{\tilde \lambda_j - \lambda_i \over 2}
   &=0\,, \\
 - \frac{\partial F}{\partial \tilde \lambda_i} =  -{i k\over 2\pi} \tilde \lambda_i - \sum_{j \neq i} \coth{\tilde \lambda_j - \tilde \lambda_i \over 2} + \sum_j \tanh{\lambda_j - \tilde \lambda_i \over 2} &= 0\,.
 }
Similar saddle-point equations appear in the context of matrix models derived from topological string theory \cite{Aganagic:2002wv, Halmagyi:2003ze}, and can be solved using powerful techniques based on holomorphy arguments and special geometry.  Such methods were used in \cite{Drukker:2010nc} to solve eqs.~\eqref{ABJMSaddle} in the limit where $N$ is taken to infinity while holding $N/k$ fixed.   Our goal in this paper is more modest than solving the matrix model for any value of the 't Hooft parameter $N/k$.  We will work at fixed $k$ and take the large $N/k$ limit.  As we will see shortly, in such a limit we can find the eigenvalue distribution using a more elementary approach.

\subsection{A Numerical Solution}
\label{NUMERICS}

To gain intuition, one can start by solving the saddle-point equations \eqref{ABJMSaddle} numerically for any values of $N$ and $k$.   One of the simplest ways to do so is to view equations \eqref{ABJMSaddle} as describing the equilibrium configuration of $2N$ point particles whose 2-d coordinates are given by the complex numbers $\lambda_j$ and $\tilde \lambda_j$ and that interact with the forces given by eq.~\eqref{ABJMSaddle}.  This equilibrium configuration can be found by introducing a time dimension and writing down equations of motion for $\lambda_j(t)$ and $\tilde \lambda_j(t)$ whose solution approaches the equilibrium configuration \eqref{ABJMSaddle} at late times in the same way as the solution to the heat equation approaches a solution to the Laplace equation at late times.  The equations of motion for the eigenvalues are
\be \label{EvaluesEoms}
\tau_\lambda \frac{d \lambda_i}{dt} = - \frac{\partial F}{\partial \lambda_i}  \,, \qquad
\tau_{\tilde \lambda} \frac{d \tilde \lambda_i}{dt} = - \frac{\partial F}{\partial \tilde \lambda_i}  \,,
\ee
where $\tau_\lambda$ and $\tau_{\tilde \lambda}$ are complex numbers that need to be chosen in such a way that the saddle point we wish to find is an attractive fixed point as $t \to \infty$.

\begin {figure} [htb]
  \center\includegraphics [width=0.6\textwidth] {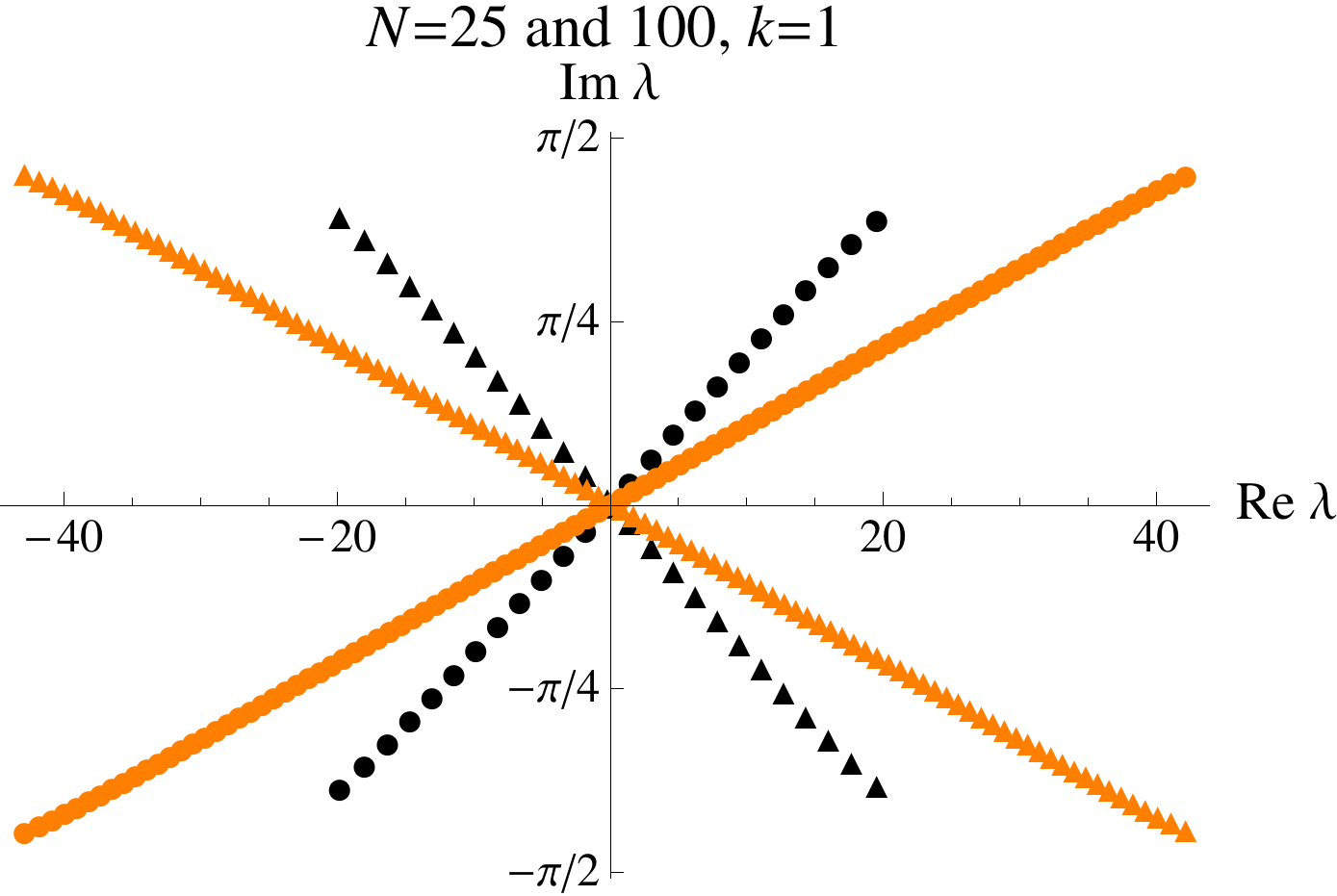}
  \caption {Numerical saddle points for the ABJM matrix model. The eigenvalues for $N=25$ are plotted in black and those for $N=100$ are plotted in orange. The plot has been obtained with $\tau_\lambda = \tau_{\tilde \lambda} = 1$. As mentioned in the text, the real parts of the eigenvalues grow with $\sqrt N$.\label{EvalsABJM}}
\end {figure}
In figure~\ref{EvalsABJM} we show typical eigenvalue distributions that can be found using the method we just explained.  There are several features of the saddle-point configurations that are worth emphasizing:
 \begin{itemize}
  \item  The eigenvalues $\lambda_j$ and $\tilde \lambda_j$ that solve \eqref{ABJMSaddle} are not real.

That the eigenvalue distributions do not lie on the real axis might be a bit puzzling given that $\lambda_i$ and $\tilde \lambda_i$ are supposed to be eigenvalues of Hermitian matrices.  However, it is well known that in general, when using the saddle-point approximation, the main contribution to an oscillatory integral may come from saddles that are not on the original integration contour but through which the integration contour can be made to pass.  We will assume that the integration contour that should be chosen in writing down the integral in eq.~\eqref{ZABJM} can be deformed so that saddle points like those in figure~\ref{EvalsABJM} are the only important ones.

   \item  The eigenvalue distributions are invariant under $\lambda_i \to -\lambda_i$ and $\tilde \lambda_i \to -\tilde \lambda_i$.

   Indeed, the saddle-point equations \eqref{ABJMSaddle} are invariant under these transformations, so it is reasonable to expect that there should be solutions that are also invariant.

  \item  In the equilibrium configuration the two types of eigenvalues are complex conjugates of each other: $\tilde \lambda_j = \bar \lambda_j$.

  Indeed, it is not hard to see that upon setting $\tilde \lambda_j = \bar \lambda_j$ the two equations in \eqref{ABJMSaddle} become equivalent, so it is consistent to look for solutions that have this property.

  \item As one increases $N$ at fixed $k$, the imaginary part of the eigenvalues stays bounded between $-\pi/2$ and $\pi/2$, while the real part grows with $N$.  We will show shortly that, for the saddle points we find, the real part grows as $N^{1/2}$ as $N\to\infty$.

  \end{itemize}

\subsection{Large $N$ Analytical Approximation}

Let us now find analytically the solution to the saddle-point equations \eqref{ABJMSaddle} in the large $N$ limit.  As explained above, we can assume $\tilde \lambda_j = \bar \lambda_j$ and write\footnote {After completing this work, we became aware that ref.~\cite{Suyama:2009pd} employs a similar ansatz.}
 \es{EvaluesReIm}{
  \lambda_j = N^{\alpha} x_j + i y_j \,,  \qquad \tilde \lambda_j = N^{\alpha} x_j - i y_j \,,
 }
where we introduced a factor of $N^{\alpha}$ multiplying the real part because we want $x_j$ and $y_j$ to be of order $O(N^0)$ and become dense in the large $N$ limit.   The constant $\alpha$ is so far arbitrary but will be determined later.

In passing to the continuum limit, we define the functions $x, y: [0, 1] \to \R$ so that
 \es{ContinuumLimit}{
  x_j = x(j/N) \,, \qquad y_j = y(j/N) \,.
 }
Let us assume we order the eigenvalues in such a way that $x$ is a strictly increasing function on $[0, 1]$.  Introducing the density of the real part of the eigenvalues
 \es{rhoDef}{
  \rho(x) = \frac{ds}{dx} \,,
 }
one can approximate \eqref{FreeEnergy} as (see Appendix \ref{app:free})
 \es{FreeFunctional}{
  F = {k \over \pi} N^{1+\alpha} \int dx\, x \rho(x) y(x) + N^{2-\alpha}\int dx\, \rho(x)^2 f(2y(x))
   + \dotsb \,,
 }
where the function $f$ is
 \es{fDef}{
  f(t) = \pi^2 - \left( \arg e^{i t} \right)^2 \,.
 }
In other words, $f$ is a periodic function with period $\pi$ given by
 \es{fABJM}{
  f(t) = \pi^2 - t^2 \qquad \text{when} \quad -\pi \leq t \leq \pi \,.
 }

It may be a little puzzling that while the discrete expression for the free energy in eq.~\eqref{FreeEnergy} is nonlocal, in the sense that there are long-range forces between the eigenvalues, its large $N$ limit \eqref{FreeFunctional} is manifestly local.  One can understand this major simplification from examining, for instance, the first saddle-point equation in \eqref{ABJMSaddle}.  The force felt by $\lambda_i$ due to interactions with far-away eigenvalues $\lambda_j$ and $\tilde \lambda_j$ is
 \es{FarForces}{
  -\coth \frac{\lambda_j - \lambda_i}{2}
   + \tanh \frac{\tilde \lambda_j - \lambda_i}{2}
  \approx -\sgn (\Re \lambda_j - \Re \lambda_i)
   + \sgn (\Re\tilde \lambda_j - \Re \lambda_i) \,,
 }
the corrections to this formula being exponentially suppressed in $\Re \lambda_j - \Re \lambda_i$ and $\Re\tilde \lambda_j - \Re \lambda_i$.  In other words, the nonlocal part of the interaction force between eigenvalues is given just by the right-hand side of eq.~\eqref{FarForces}.  The nonlocal part of the force vanishes when $\Re \lambda_j = \Re \tilde \lambda_j$, so in assuming that the two eigenvalue distributions are complex conjugates of each other, we effectively arranged for an exact cancellation of nonlocal effects.  All that is left are short-range forces, which in the large $N$ limit are described by the local action \eqref{FreeFunctional}.

One can view $F$ as a functional of $\rho(x)$ and $y(x)$ and look for its saddle points in the set
 \es{rhoSet}{
  {\cal C} = \left\{ (\rho, y): \int dx\, \rho(x) = 1; \rho(x) \geq 0 \text{ pointwise} \right\} \,.
 }
These constraints mean that $\rho$ is a normalized density.   Motivated by the numerical analysis we performed, we assume that $\rho$ and $y$ describe a connected distribution of eigenvalues contained in a bounded region of the complex plane.

Let us assume a saddle point for $F$ exists.  As $N\to \infty$, we need the two terms in \eqref{FreeFunctional} to be of the same order in $N$ in order to have nontrivial solutions, so from now on we will set
 \es{GotAlpha}{
  \alpha = \frac 12\,.
 }
The real part of the eigenvalues therefore grows as $N^{1/2}$, and to leading order, the free energy behaves as $N^{3/2}$ at large $N$.
In writing \eqref{FreeFunctional} we omitted the last two terms from eq.~\eqref{FreeEnergy}. They do not depend on $\rho$ or $y$ and hence do not affect the saddle-point equations.  They are also lower order in $N$ given the choice of $\alpha$.

To find a saddle point for $F$, one can add a Lagrange multiplier $\mu$ to \eqref{FreeFunctional} and extremize
 \es{tildeF}{
  \tilde F = N^{3/2} \left[ {k \over \pi}  \int dx\, x \rho(x) y(x) + \int dx\, \rho(x)^2 f(2y(x))
   - {\mu \over 2 \pi}  \left(\int dx\, \rho(x) - 1 \right) \right]
 }
instead of \eqref{FreeFunctional}.  As long as $\rho(x) > 0$, the saddle-point eigenvalue distribution satisfies the equations
 \es{eoms}{
 4 \pi  \rho(x) f(2y(x))&=  \mu - 2 k x y(x) \,, \\
 2 \pi  \rho(x) f'(2y(x)) &= - kx  \,.
 }

Plugging \eqref{fABJM} into \eqref{eoms} one obtains
 \es{Gotrhob}{
  \rho(x) = {\mu \over 4 \pi^3} \,, \qquad y(x) = {\pi^2 k x \over 2 \mu}\,,
 }
as long as $-\pi/2 \leq y(x) \leq \pi/2$.  If $\rho$ is supported on
$[-x_{*}, x_{*}]$ for some $x_{*}>0$ that we will determine shortly, we can calculate $\mu$ from the normalization of the density $\rho(x)$:
 \es{rhoNormalizationImpliesmu}{
  \int_{-x_{*}}^{x_{*}}dx\, \rho(x) = 1  \qquad
   \Longrightarrow \qquad  \mu = {2 \pi^3 \over x_{*} }  \,.
 }
Plugging this formula back into \eqref{FreeFunctional}, we obtain the free energy in terms of $x_{*}$:
 \es{Freeamax}{
  F = {N^{3/2} (12 \pi^4 + k^2 x_{*}^4) \over 24 \pi^2 x_{*}} + o(N^{3/2}) \,.
 }
This expression is extremized when
 \es{Gotamax}{
  x_{*} = \pi \sqrt{2 \over k} \,, \qquad y(x_{*}) = \frac {\pi}2 \,.
 }
Luckily, the answer $y(x_{*}) = \pi/2$ is consistent with our assumption that $-\pi/2 \leq y(x) \leq \pi/2$ without which eq.~\eqref{Gotrhob} would not be correct.  In Appendix~\ref{MOREDETAILS} we check that assuming $y(x_{*})>\pi/2$ implies a contradiction.  The extremum of $F$ obtained from eqs.~\eqref{Freeamax} and \eqref{Gotamax} is
 \es{GotFmin}{
    F = {\pi \sqrt{2} \over 3} k^{1/2} N^{3/2} + o(N^{3/2}) \,.
 }
 This result agrees with the free energy found in \cite{Drukker:2010nc} using the regularized Euclidean action in $AdS_4$ \cite{Emparan:1999pm}.
\begin {figure} [htb]
  \center\includegraphics [width=0.6\textwidth] {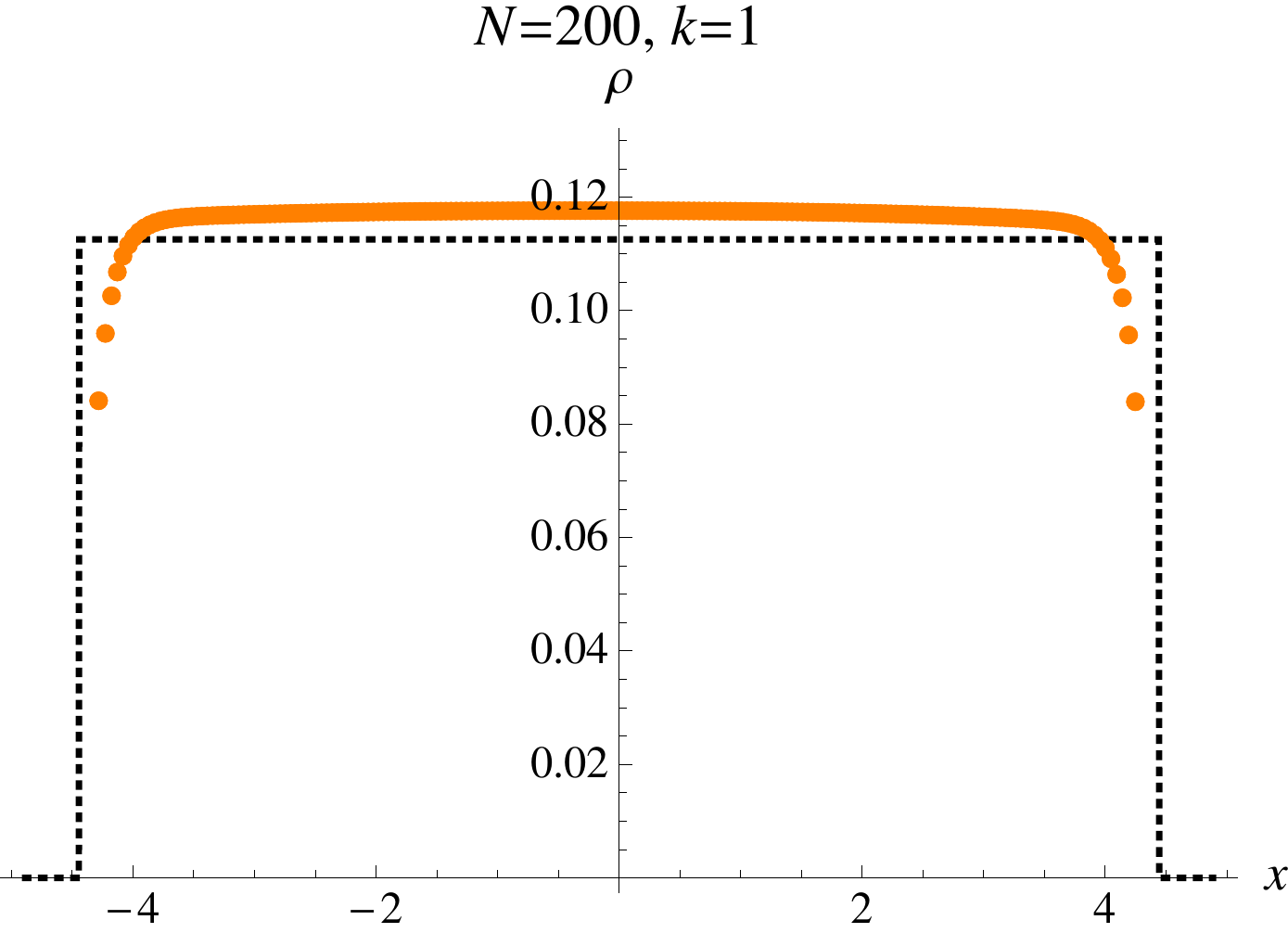}
  \caption {Comparison between analytical prediction and numerical results for the density of eigenvalues $\rho$ defined in eq.~\eqref {rhoDef}. The dotted black line represents the analytical calculation, and the numerical result is shown in orange dots.\label {DensityABJM}}
\end {figure}%

In the large $N$ limit the eigenvalues therefore condense on two line segments, and on these two line segments they have uniform density.  In figure~\ref{DensityABJM} we compare the analytical result for the density with the numerical one.

We would like to compare the location of our eigenvalue distributions
with the results of~\cite{Drukker:2010nc}.
Noting a similarity between the ABJM matrix model
and the $S^3/{\mathbb Z}_2$ matrix model solved in~\cite{Halmagyi:2003ze}, Drukker, Marino, and Putrov \cite{Drukker:2010nc} wrote down a resolvent for the ABJM model.  This resolvent has cuts in the $\lambda$ plane corresponding to the locations of the eigenvalues.  In particular, it has a cut where the $\lambda_i$ eigenvalues are located and a second cut where the $\tilde \lambda_i$
eigenvalues are located but shifted by $\pi i$.
More specifically, the resolvent has the form
\be
\omega(\lambda) = 2 \log \left(\frac{1}{2} \left[
\sqrt{(e^{\lambda}+b)(e^{\lambda}+1/b)} - \sqrt{(e^{\lambda}-a)(e^\lambda-1/a)} \right] \right) \,.
\ee
where $a + 1/a + b + 1/b = 4$
and at strong coupling,
\be
a + \frac{1}{a} - b - \frac{1}{b} = 2 i \exp \left( \pi \sqrt{\frac{2N}{k} - \frac{1}{12}} \right) + \ldots
\ee
The ellipses denote terms exponentially suppressed  in $N/k$ relative to the leading term.
Solving the equations for $a$ and $b$, we find that the branch points in the $\lambda$ plane are at
\be
\pm \log a =\pi \sqrt{\frac{2N}{k} - \frac{1}{12}}+   \frac{i \pi}{2} \,,
\ee
\be
\pm \log b =-\pi \sqrt{\frac{2N}{k}- \frac{1}{12}} +  \frac{i \pi}{2} \,.
\ee
These expressions are in agreement with (\ref{Gotamax}) in the large $N$ limit.

Let us also try to compare our findings with the exact results found for the supersymmetric Wilson loops in ABJM theory~\cite{Drukker:2010nc, Marino:2009jd}.  The expectation values of $1/6$ and $1/2$ supersymmetric Wilson loops are proportional, respectively, to the expectation values of $\sum_{i=1}^N e^{\lambda_i}$ and  $\sum_{i=1}^N \left[e^{\lambda_i}+ e^{\tilde \lambda_i}\right]$ in the matrix model~\cite{Drukker:2010nc, Drukker:2009hy, Marino:2009jd, Kapustin:2009kz}.  In our approach, these quantities become
\begin{align}
\langle W_{\square}^{1/6} \rangle
&= \frac{2 \pi i N}{k} \int_{-x_*}^{x_*} e^{\lambda(x)} \rho(x) dx \,, \label{OneSixth}\\
\langle W_{\square}^{1/2} \rangle
&=  \frac{2 \pi i N}{k} \int_{-x_*}^{x_*} \left( e^{\lambda(x)}  + e^{\tilde \lambda(x)} \right) \rho(x) dx \,. \label{OneHalf}
\end {align}
If we evaluate \eqref{OneSixth} and \eqref{OneHalf} using the saddle point we have found, we get
\begin{align}
\langle W_{\square}^{1/6} \rangle  &\approx   - \sqrt{\frac{N}{2k}} e^{ \pi \sqrt{2N/k}} \,, \\
\langle W_{\square}^{1/2} \rangle  &\approx \frac{i}{2} e^{ \pi \sqrt{2N/k}} \,.
\end{align}
The exponents in these formulae agree with the results in~\cite{Drukker:2010nc, Drukker:2009hy, Marino:2009jd}.

We should keep in mind that the ABJM model has a type IIA string interpretation only in the limit where $N/k \gg 1$, $N^{1/2}/k^{5/2} \ll 1$. These conditions apply only in the limit where both $N$ and $k$ are taken to infinity.  Our approximations are only applicable in the M-theory limit where $N$ is taken to infinity at fixed $k$. Thus our Wilson loops have a dual interpretation as wrapped M2-branes in M-theory rather than as strings in type IIA string theory.

\section{Necklace Quiver Gauge Theories}

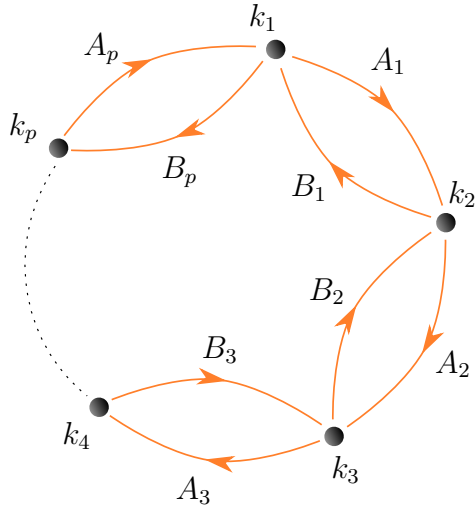
\begin {figure} [!t]
  \centering
  \newcommand {\svgwidth} {0.4\textwidth}

\begingroup
  \makeatletter
  \providecommand\color[2][]{%
    \errmessage{(Inkscape) Color is used for the text in Inkscape, but the package 'color.sty' is not loaded}
    \renewcommand\color[2][]{}%
  }
  \providecommand\transparent[1]{%
    \errmessage{(Inkscape) Transparency is used (non-zero) for the text in Inkscape, but the package 'transparent.sty' is not loaded}
    \renewcommand\transparent[1]{}%
  }
  \providecommand\rotatebox[2]{#2}
  \ifx\svgwidth\undefined
    \setlength{\unitlength}{350.33000488pt}
  \else
    \setlength{\unitlength}{\svgwidth}
  \fi
  \global\let\svgwidth\undefined
  \makeatother
  \begin{picture}(1,0.99670505)%
    \put(0,0){\includegraphics[width=\unitlength]{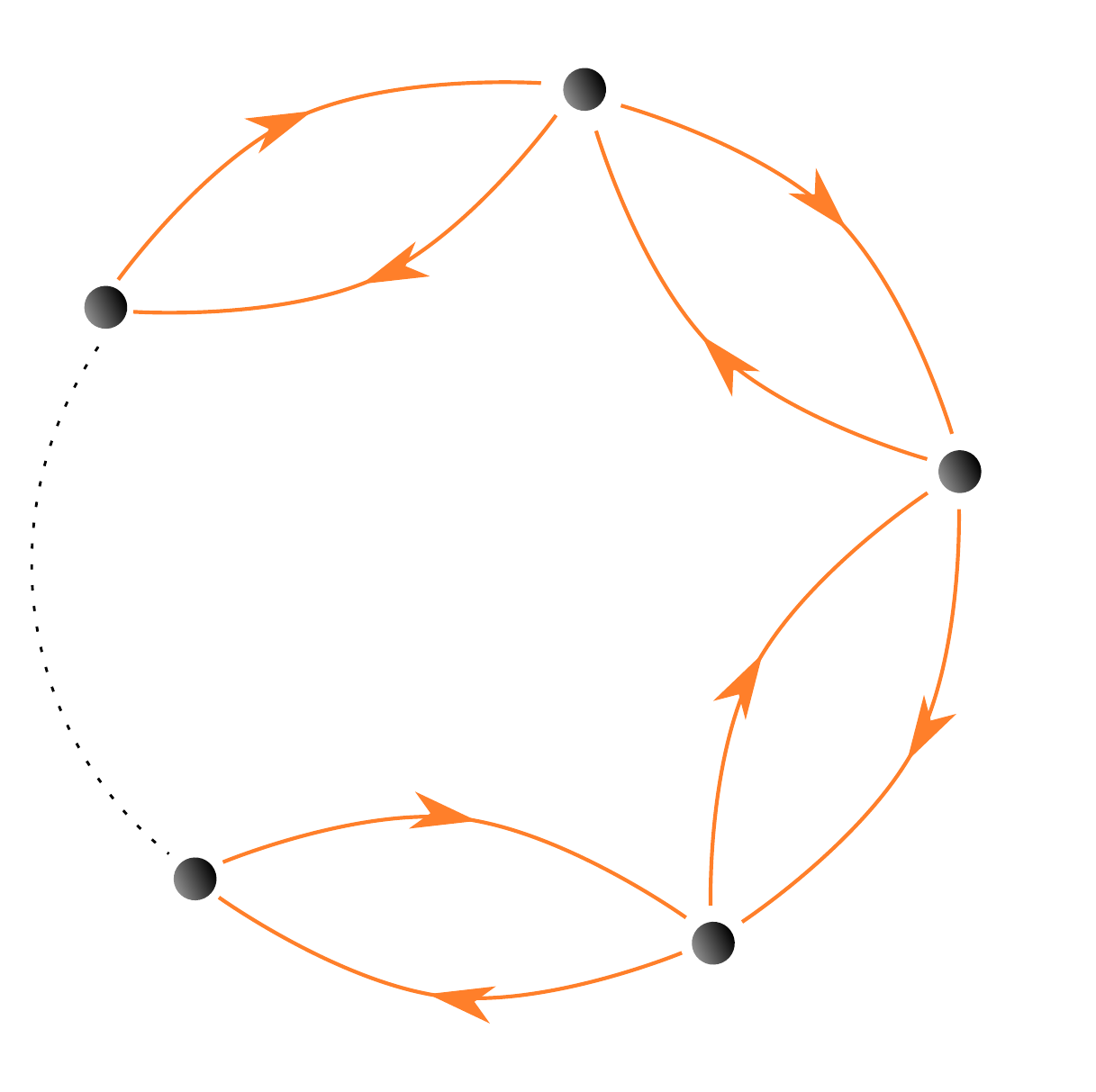}}%
    \put(0.47931781,1.00641416){\color[rgb]{0,0,0}\makebox(0,0)[lt]{\begin{minipage}{0.23814279\unitlength}\raggedright $k_1$\end{minipage}}}%
    \put(0.88049059,0.65083102){\color[rgb]{0,0,0}\makebox(0,0)[lt]{\begin{minipage}{0.13375145\unitlength}\raggedright $k_2$\end{minipage}}}%
    \put(0.64573236,0.09625194){\color[rgb]{0,0,0}\makebox(0,0)[lt]{\begin{minipage}{0.16963595\unitlength}\raggedright $k_3$\end{minipage}}}%
    \put(0.10414096,0.1647588){\color[rgb]{0,0,0}\makebox(0,0)[lt]{\begin{minipage}{0.17616049\unitlength}\raggedright $k_4$\end{minipage}}}%
    \put(-0.00339161,0.78784474){\color[rgb]{0,0,0}\makebox(0,0)[lt]{\begin{minipage}{0.19573384\unitlength}\raggedright $k_p$\\ \end{minipage}}}%
    \put(0.14680037,0.90354507){\color[rgb]{0,0,0}\makebox(0,0)[lb]{\smash{$A_p$}}}%
    \put(0.72421507,0.87418506){\color[rgb]{0,0,0}\makebox(0,0)[lb]{\smash{$A_1$}}}%
    \put(0.85470431,0.26414798){\color[rgb]{0,0,0}\makebox(0,0)[lb]{\smash{$A_2$}}}%
    \put(0.33274746,0.00969401){\color[rgb]{0,0,0}\makebox(0,0)[lb]{\smash{$A_3$}}}%
    \put(0.56436585,0.61973106){\color[rgb]{0,0,0}\makebox(0,0)[lb]{\smash{$B_1$}}}%
    \put(0.60025034,0.41094828){\color[rgb]{0,0,0}\makebox(0,0)[lb]{\smash{$B_2$}}}%
    \put(0.38494317,0.29350807){\color[rgb]{0,0,0}\makebox(0,0)[lb]{\smash{$B_3$}}}%
    \put(0.30338738,0.65561554){\color[rgb]{0,0,0}\makebox(0,0)[lb]{\smash{$B_p$}}}%
  \end{picture}%
\endgroup

  \caption {Necklace quiver diagrams for $U(N)^p$ Chern-Simons gauge theories.\label {NecklaceQuiver}}
\end {figure}
In this section we consider a class of quiver Chern-Simons $U(N)_{k_1}\times U(N)_{k_2}\times \dotsm \times U(N)_{k_p}$ gauge theories whose quiver diagrams look like necklaces (see figure~\ref{NecklaceQuiver}) \cite{Imamura:2008nn, Jafferis:2008qz}. In the ${\cal N}=2$ superspace formulation the theory is coupled to bifundamental chiral superfields $A_a$ and $B_a$, $a=1,2,\ldots, p$, whose interactions are governed by the quartic superpotential
\es{quartsup}
{W =-\sum_{a=1}^p {2\pi \over k_a} ( B_{a-1} A_{a-1} - A_a B_a)^2 \,.}
The 4-d ``parent theories'' for these Chern-Simons models, i.e.~the 4-d gauge theories with the same type of quiver diagrams and superpotentials, arise on $N$ D3-branes placed on generalized conifolds \cite{Klebanov:1998hh,Gubser:1998ia}.  For $\sum_{a=1}^p k_a=0$ and $N=1$ the moduli space of each 3-d model was calculated in \cite{Imamura:2008nn, Jafferis:2008qz} and shown to be given by a certain hyper-K\"ahler cone in four complex dimensions \cite{Boyer:1993,Boyer:1998sf,Boyer:1994}. Therefore, it was conjectured that such a model describes the low-energy dynamics of $N$ coincident M2-branes placed at the tip of this cone. The quiver Chern-Simons gauge theories were therefore conjectured to be dual to the $AdS_4\times Y$ backgrounds of M-theory where $Y$ are the bases of the cones. These gauge theories are natural generalizations of the ABJM model that is found for $p=2$.

The gauge theories in figure~\ref{NecklaceQuiver} arise from the following type IIB brane construction \cite{Imamura:2008nn}. We consider $N$ D3-branes filling directions $0123$ where
$x^3$ is a circle, and then add a sequence of $(1,q)$ 5-branes (the bound states of an NS5-brane and $q_a$ D5-branes) filling directions $012$ and located at specific points along the $x^3$ circle. If we express the CS levels as
 \es{kToq}{
  k_a = q_{a+1} - q_a \,,
 }
where $q_a$ are integers, then as we move around the circle we find the ordering $(1, q_a)$ of the
5-branes. This picture is helpful in considering certain transformations in the gauge theory that are analogous to the Seiberg duality in four-dimensional gauge theory \cite{Aharony:2008gk,Giveon:2008zn, Amariti:2009rb}. As in that case, these transformations are related to interchange of adjacent branes and thus correspond to interchange of $q_a$ and $q_{a+1}$.  (There is also a shift in the rank of one of the gauge groups that may be neglected in the large $N$ limit.)
Our explicit answer for the free energy will have this symmetry.

For a general set of Chern-Simons levels such a $p>2$ gauge theory has ${\cal N}=3$ superconformal invariance, but in the special case where $p$ is even and the CS levels are $(k, -k, k, -k, \ldots)$ the supersymmetry is enhanced to ${\cal N}=4$ \cite{Benna:2008zy, Imamura:2008nn, Hosomichi:2008jd}. Then $Y=S^7/(\Z_{p/2}\times \Z_{kp/2})$ and $\Vol (Y)= 4\pi^4 / (3kp^2)$. For more general $k_i$ the eight-dimensional cone is not an orbifold, which complicates the calculation of its volume. Nevertheless, these volumes were computed in \cite{Lee:2006ys,Yee:2006ba}, and we can compare the result on the gravity side with our calculation of the free energy.

\subsection{Multi-Matrix Models}

As explained in \cite{Kapustin:2009kz}, the partition function for the necklace quivers in figure~\ref{NecklaceQuiver} localizes on configurations where the scalars $\sigma_a$ in the ${\cal N} = 2$ vector multiplets are constant Hermitian matrices.  Denoting by $\lambda_{a, i}$, $1 \leq i \leq N$, the eigenvalues of $\sigma_a$, the partition function takes the form of the matrix integral
 \es{NecklaceModel}{
   Z =  \frac{1}{(N!)^p}\int \left(\prod_{a,i} \frac{d \lambda_{a,i} }{2 \pi} \right)
      \prod_{a=1}^p \left( \frac{ \prod_{i<j}
     \left(2  \sinh \frac{\lambda_{a,i} - \lambda_{a,j}}{2}  \right)^2}
  {  \prod_{i,j} 2 \cosh \frac{\lambda_{a,i} - \lambda_{a+1,j}}{2} }
   \exp \left[\frac{i}{4 \pi}  \sum_{i}  k_a \lambda_{a,i}^2  \right]
    \right)  \,.
}
The normalization of the partition function was chosen so that it agrees with the ABJM result from eq.~\eqref{ZABJM} in the case $p=2$.  As in the ABJM case, the integration contour should be taken to be the real axis in each integration variable.   The saddle-point equations following from \eqref{NecklaceModel} are
 \es{NecklaceSaddle}{
  {i k_a\over \pi} \lambda_{a, i} - 2 \sum_{j \neq i} \coth{\lambda_{a, j} - \lambda_{a, i} \over 2}
   +  \sum_j \tanh{\lambda_{a+1, j} - \lambda_{a, i} \over 2}
   +  \sum_j \tanh{\lambda_{a-1, j} - \lambda_{a, i} \over 2}
   =0\,.
 }
These equations can be solved numerically using the method described in section~\ref{NUMERICS}:  By replacing the right-hand side of these equations by $\tau_a d\lambda_{a, j} / dt$, we obtain a system of first order differential equations whose solution converges at late times $t$ to a solution of eq.~\eqref{NecklaceSaddle} provided that the constants $\tau_a$ are chosen appropriately.  We will now show how to obtain an approximate analytical solution valid in the limit where $N$ is taken to be large and $k$ is held fixed.

Based on our intuition from the ABJM model, let us assume that in this case too the real part of the eigenvalues behaves as $N^{1/2}$ at large $N$ while the imaginary part is of order one.  So if one writes
 \es{lambdaLargeN}{
   \lambda_{a, j} = N^{1/2} x_{a, j} + i y_{a, j}  \,,
 }
then the quantities $x_{a, j}$ and $y_{a, j}$ become dense in the large $N$ limit.  Under this assumption, we will be able to solve the saddle-point equations to leading order in $N$ in a self-consistent way.  We can pass to the continuum limit by considering the normalized
densities $\rho_a(x)$ of the $x_{a, j}$ together with the continuous functions $y_a(x)$ that describe the imaginary parts of the eigenvalues as functions of $x$.  Let us first make a rough approximation to the saddle-point equations \eqref{NecklaceSaddle}.  When $N$ is large, we have
 \es{RoughApproximation}{
  \coth{\lambda_{a, j} - \lambda_{a, i} \over 2} \approx \sgn \left( x_{a, j} -  x_{a, i} \right) \,,
   \qquad
   \tanh{\lambda_{a, j} - \lambda_{a\pm 1, i} \over 2} \approx \sgn \left( x_{a, j} -  x_{a\pm1, i} \right) \,.
 }
To leading order in $N$, the saddle-point equations then become
 \es{SaddleDensity}{
  \int dx' \left[ 2 \rho_a(x') - \rho_{a+1}(x') - \rho_{a-1}(x') \right] \sgn(x'-x) = 0 \,.
 }
Differentiating with respect to $x$, we immediately conclude that all $\rho_a$ must be equal to one another to leading order in $N$, so we can write $\rho_a(x) \equiv \rho(x)$ for some density function $\rho(x)$ that is normalized as
 \es{rhoNormalization}{
  \int dx\, \rho(x)  = 1 \,.
 }

With the simplifying assumption that the densities $\rho_a$ are equal, one can go back to the integral \eqref{NecklaceModel} and calculate the free energy functional $F[\rho, y_a]$ to leading order in $N$ (see Appendix \ref{app:free}):
\es{FNecklace}{
  F[\rho, y_a] = \frac{N^{3/2}}{2\pi}  \int dx\, &x \rho(x) \sum_{a=1}^p
     k_a y_a(x)  \\
     &+ \frac{N^{3/2}}2 \int dx\, \rho(x)^2 \sum_{a=1}^p f( y_{a+1}(x) - y_a(x))
     + o(N^{3/2})  \,,
 }
where $f$ is the same function that was defined in \eqref{fDef}.  We wish to evaluate the integral \eqref{NecklaceModel} in the saddle-point approximation where it equals $Z = e^{-F}$, the free energy $F$ being an appropriate critical point of $F[\rho, y_a]$.  Let us assume that the eigenvalue distribution corresponding to this saddle point is connected, symmetric about $x = y = 0$, and bounded.

In looking for the eigenvalue distribution that extremizes \eqref{FNecklace} to order $O(N^{3/2})$, an important observation is that, in fact, one cannot find this distribution, because to this order in $N$ $F[\rho, y_a]$ has a flat direction given by $y_a(x) \to y_a(x) + \delta y(x)$ for any function $\delta y(x)$.  The second term in eq.~\eqref{FNecklace} is clearly invariant under this transformation, and the first term is also invariant because $\sum_{a=1}^p k_a = 0$.  The existence of this flat direction is not a problem at all if one just wants to compute the free energy $F$ to leading order in $N$.  If one's goal is instead to find the eigenvalue distributions for the saddle point, subleading corrections to \eqref{FNecklace} that presumably lift this flat direction must be taken into account.  In this paper we will content ourselves with calculating the free energy to order $O(N^{3/2})$, and will leave a careful analysis of how the flat direction gets lifted for future work.

Before we examine the extremization of the free energy functional \eqref{FNecklace} in more detail, let us make a few comments and present a result that follows already from the discussion above.  Suppose we manage to find a saddle point of $F$ by extremizing \eqref{FNecklace} for a quiver Chern-Simons gauge theory that in the large $N$ limit and at strong 't Hooft coupling is dual to an $AdS_4 \times Y$ M-theory background.   Let us assume that this saddle point gives the most important contribution to the partition function.  What can we learn?  From \eqref{FNecklace} one may infer that the free energy grows as $N^{3/2}$ at large $N$ as expected from supergravity, so our computation provides a gauge theory explanation of this $N^{3/2}$ behavior.  Moreover, one can compare the free energy we obtain with the exact M-theory result
 \es{FExpectation}{
  F =  N^{3/2} \sqrt{\frac{2 \pi^6} {27\Vol(Y)}}
 }
that can be derived as a straightforward generalization of the ABJM computation in \cite{Drukker:2010nc}. Via this formula we will compare successfully our matrix model results with the expressions for the volumes of tri-Sasaki Einstein space available in the literature \cite{Lee:2006ys,Yee:2006ba}.

\subsection{A Class of Orbifold Chern-Simons Theories}

The vacuum moduli space of the nonchiral quivers with alternating CS levels $(k, -k, k, -k, \ldots)$ and $N=1$ is the orbifold $\C^4 / \left( \Z_{p/2} \times \Z_{kp/2} \right)$ \cite{Imamura:2008nn}.  There is an induced orbifold action on the unit 7-sphere in $\C^4$, and thus the internal space $Y$ is $S^7 / \left( \Z_{p/2} \times \Z_{kp/2} \right)$. Consequently, we expect
 \es{VolYOrbifold}{
  \Vol(Y) = \frac{4 \Vol(S^7)}{kp^2} = \frac{4 \pi^4}{3 k p^2}\,,
 }
where in the second equality we used the round 7-sphere volume $\Vol(S^7) = \pi^4/3$.

This formula can be reproduced very easily from the matrix model computation. The saddle-point equations \eqref{NecklaceSaddle} are solved by setting $\lambda_{2a, i} = \lambda_i$ and $\lambda_{2a+1, i} = \tilde \lambda_i$, $\lambda_i$ and $\tilde \lambda_i$ being the eigenvalues for the saddle point of the ABJM matrix model discussed in detail in section~\ref{ABJM}.  The free energy of the $p$-node quiver with CS levels $(k, -k, k, -k, \ldots)$ is therefore $p/2$ times the free energy in the ABJM model, and thus
 \es{GotFOrbifold}{
  F =  \frac p2 F_{\rm ABJM} = {\pi \sqrt{2} \over 6} p k^{1/2} N^{3/2} + o(N^{3/2}) \,.
 }
Using eq.~\eqref{FExpectation}, one immediately reproduces the volume of the $S^7$ orbifold in eq.~\eqref{VolYOrbifold}.

\subsection{Warm-up:  A Four-Node Quiver}

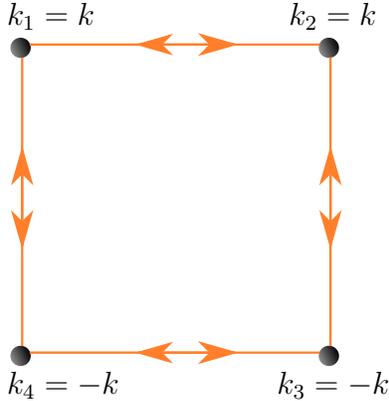
\begin {figure} [thb]
  \centering
  \newcommand {\svgwidth} {0.3\textwidth}

\begingroup
  \makeatletter
  \providecommand\color[2][]{%
    \errmessage{(Inkscape) Color is used for the text in Inkscape, but the package 'color.sty' is not loaded}
    \renewcommand\color[2][]{}%
  }
  \providecommand\transparent[1]{%
    \errmessage{(Inkscape) Transparency is used (non-zero) for the text in Inkscape, but the package 'transparent.sty' is not loaded}
    \renewcommand\transparent[1]{}%
  }
  \providecommand\rotatebox[2]{#2}
  \ifx\svgwidth\undefined
    \setlength{\unitlength}{251.36621094pt}
  \else
    \setlength{\unitlength}{\svgwidth}
  \fi
  \global\let\svgwidth\undefined
  \makeatother
  \begin{picture}(1,1.04819007)%
    \put(0,0){\includegraphics[width=\unitlength]{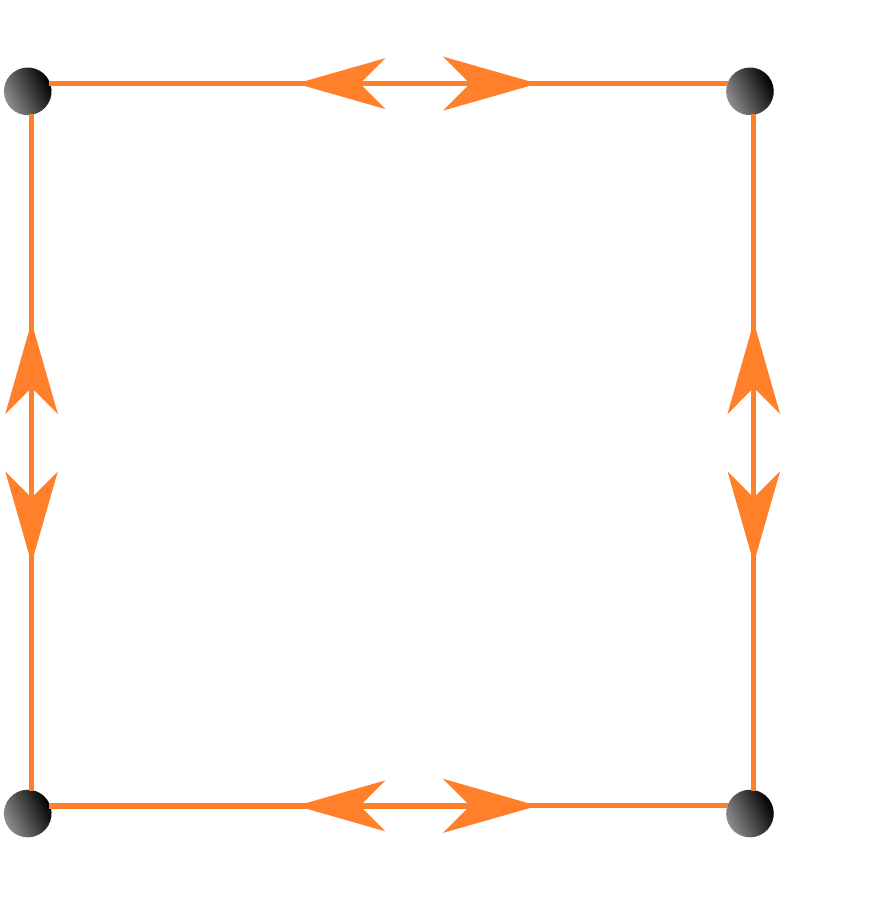}}%
    \put(-0.00475524,1.00466543){\color[rgb]{0,0,0}\makebox(0,0)[lb]{\smash{$k_1=k$}}}%
    \put(0.75225074,1.00466541){\color[rgb]{0,0,0}\makebox(0,0)[lb]{\smash{$k_2=k$}}}%
    \put(-0.00475526,0.01351053){\color[rgb]{0,0,0}\makebox(0,0)[lb]{\smash{$k_4=-k$}}}%
    \put(0.72042466,0.01351053){\color[rgb]{0,0,0}\makebox(0,0)[lb]{\smash{$k_3=-k$}}}%
  \end{picture}%
\endgroup

  \caption {Four-node quiver diagram obtained as a particular case of the general quivers presented in figure~\ref{NecklaceQuiver}.\label {SquareQuiver}}
\end {figure}
Another case we can easily solve using the approximation scheme developed above is that of the four-node quiver with CS levels $k_a = (k, k, -k, -k)$ (see figure~\ref{SquareQuiver}).  The two $\Z_2$ symmetries of the quiver, one acting by interchanging nodes $1 \leftrightarrow 4$ and $2 \leftrightarrow 3$ and the other by interchanging nodes $1 \leftrightarrow 2$ and $3\leftrightarrow 4$, allow us to set
 \es{ConsistentChoices}{
  \lambda_{1,j} = \lambda_{2,j}  = \lambda_j \,,
  \qquad \lambda_{3,j}  = \lambda_{4,j}  = \tilde \lambda_j \,.
 }
Moreover, in the saddle-point equations \eqref{NecklaceSaddle} it is consistent to set $\tilde \lambda_j = \bar \lambda_j$ as in the ABJM case, which reduces our task to finding a single eigenvalue distribution $\lambda_i$.  In passing to the continuum limit, we should therefore set
 \es{yAnsatz}{
  y_1 = y_2 = -y_3 = -y_4 = y \,.
 }
The free energy functional \eqref{FNecklace} then becomes
\es{FFourNode}{
  F[\rho, y] = \frac{2 k N^{3/2}}{\pi}  \int dx\, x \rho(x)
      y(x)
     + N^{3/2} \int dx\, \rho(x)^2 \left[ \pi^2 + f( 2 y(x)) \right]
     + o(N^{3/2})  \,.
 }

In the paragraph following eq.~\eqref{FNecklace} we discussed how for arbitrary $p$-node quivers we would not be able to solve for the $y_a$ themselves, but only for differences of consecutive $y_a$, because the leading large $N$ contribution to the free energy functional is invariant under the shifts $y_a \to y_a + \delta y$ for any function $\delta y$.  In the case of the $(k, k, -k, -k)$ quiver we will, however, be able to determine the location of the eigenvalues exactly, because the ansatz  \eqref{yAnsatz} breaks this shift symmetry.

In order to find the saddle points of \eqref{FFourNode} in the set \eqref{rhoSet}, we should add a Lagrange multiplier $\mu$ to enforce the normalization condition for $\rho$ and extremize the functional
 \es{FtildeFour}{
  \tilde F[\rho, y] = F - \frac{N^{3/2} }{2 \pi} \mu \left( \int dx\, \rho(x) -1\right) \,.
 }

Let us assume the eigenvalue distribution is symmetric around $x = y = 0$ and ranges between $[-x_*, x_*]$. Let us focus on the region where $x\geq 0$.  Solving the equations of motion we obtain
 \es{rhobFourRegion1}{
  \rho(x) = {\mu \over 8 \pi^3} \,, \qquad y(x) = {2 k \pi^2 x \over \mu}\,, \qquad
   \text{if} \quad \abs {y(x)} \leq \frac \pi2 \,.
 }
Since $\rho(x) > 0$ in this region, we have $\mu>0$ and $y(x) \geq 0$.  Assuming $y(x_*) < \pi/2$, we can find $\mu$ in terms of $x_*$ from the normalization condition for $\rho$, and then express $F$ in terms of $x_*$ and extremize it.  The extremization yields $x_* = 2^{1/4} \pi / \sqrt{k}$ and $y(x_*) = \pi/\sqrt{2} > \pi/2$, which suggests that the assumption $y(x_*) < \pi/2$ might be wrong.  One could imagine that $y(x_*) > \pi/2$, but solving the saddle-point equations in the region where $y>\pi/2$ would yield $\rho(x)<0$.

The correct answer is $y(x_*) = \pi/2$, and in fact $y(x) = \pi/2$ on some interval $[x_{\pi/2}, x_*]$ with $0<x_{\pi/2}<x_*$.  On this interval,
 \es{rhobFourRegion3}{
  \rho(x) = {\mu - 2 k \pi x \over 4 \pi^3} \,, \qquad
    \quad y(x) = {\pi \over 2}  \,,
 }
where in obtaining these equations we only varied \eqref{FtildeFour} with respect to $\rho$.  The quantity $x_{\pi/2}$ can be obtained from setting $y(x_{\pi/2}) = \pi/2$ in \eqref{rhobFourRegion1}:
 \es{Gotxstar}{
  x_{\pi/2} = \frac{\mu}{4 \pi k} \,.
 }
One can now find $\mu$ by imposing the normalization condition for $\rho$, and then express the free energy $F$ in terms of $x_*$ and extremize with respect to $x_*$.  The result is that
 \es{GotamaxFour}{
  x_* = 2 x_{\pi/2} = 2\pi \sqrt {\frac 2 {3 k}} \,,
   \qquad
   \mu = 4 \pi^2 \sqrt{\frac{2k}{3}} \,.
 }
\begin {figure} [tbh]
  \center\includegraphics [width=\textwidth] {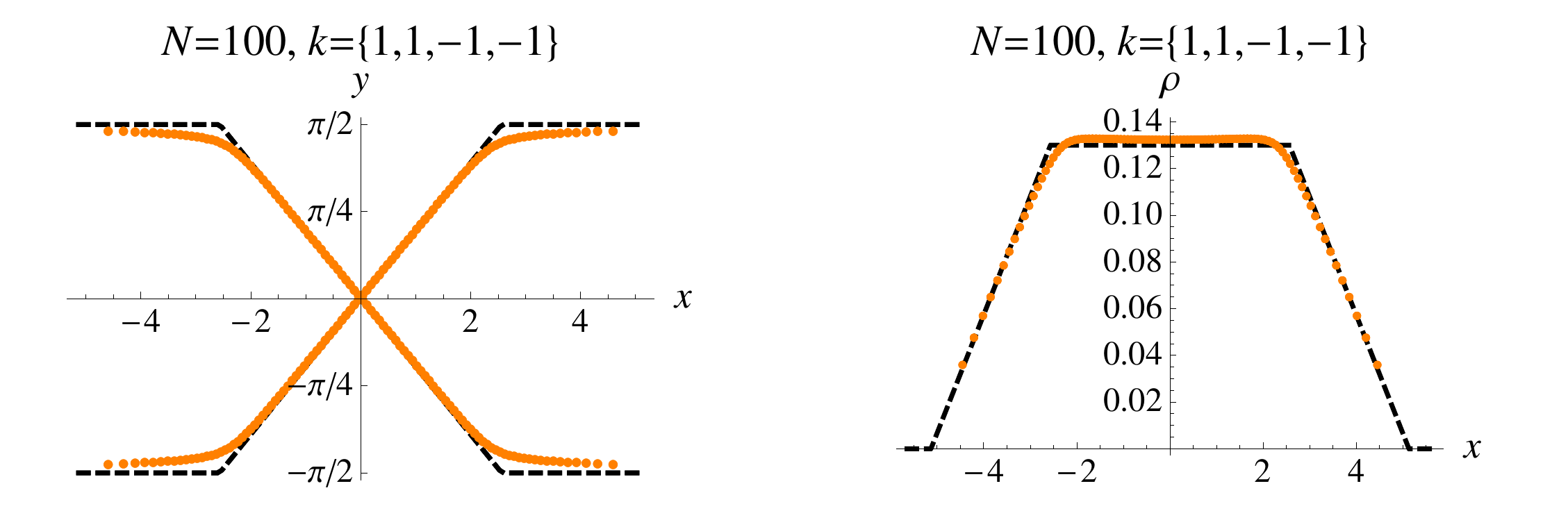}
  \caption{Comparison between numerics and analytical prediction for the four-node quiver with $k = \{1, 1, -1, -1\}$.  The dotted black lines represent the large $N$ analytical prediction, and the orange dots represent numerical results.\label {FourNodeSimple}}
\end {figure}
The density of eigenvalues is constant on $[-x_{\pi/2}, x_{\pi/2}]$ and then drops linearly to zero on $[-x_*, -x_{\pi/2}]$ and $[x_{\pi/2}, x_*]$.  See figure~\ref{FourNodeSimple} for a comparison of this analytical prediction with a numerical solution of the saddle-point equations.

The free energy for this model can be computed from \eqref{FFourNode}:
 \es{GotFFour}{
  F = \sqrt{32 \over 27} \pi k^{1/2} N^{3/2} + o(N^{3/2}) \,.
 }
Using \eqref{FExpectation}, we infer that the gravity dual of the Chern-Simons quiver gauge theory with CS levels $(k, k, -k, -k)$ is $AdS_4 \times Y$ where the volume of the compact space $Y$ is
 \es{volYFour}{
  \Vol(Y) = \frac{\pi^4}{16k}\,.
 }
 Satisfyingly, this result is in agreement with the calculation of the corresponding integral representation given in \cite{Yee:2006ba} for $k=1$, which we will review in section 4.

\begin {figure} [thb]
  \centering
  \newcommand {\svgwidth} {0.3\textwidth}

\begingroup
  \makeatletter
  \providecommand\color[2][]{%
    \errmessage{(Inkscape) Color is used for the text in Inkscape, but the package 'color.sty' is not loaded}
    \renewcommand\color[2][]{}%
  }
  \providecommand\transparent[1]{%
    \errmessage{(Inkscape) Transparency is used (non-zero) for the text in Inkscape, but the package 'transparent.sty' is not loaded}
    \renewcommand\transparent[1]{}%
  }
  \providecommand\rotatebox[2]{#2}
  \ifx\svgwidth\undefined
    \setlength{\unitlength}{251.36621094pt}
  \else
    \setlength{\unitlength}{\svgwidth}
  \fi
  \global\let\svgwidth\undefined
  \makeatother
  \begin{picture}(1,1.04819007)%
    \put(0,0){\includegraphics[width=\unitlength]{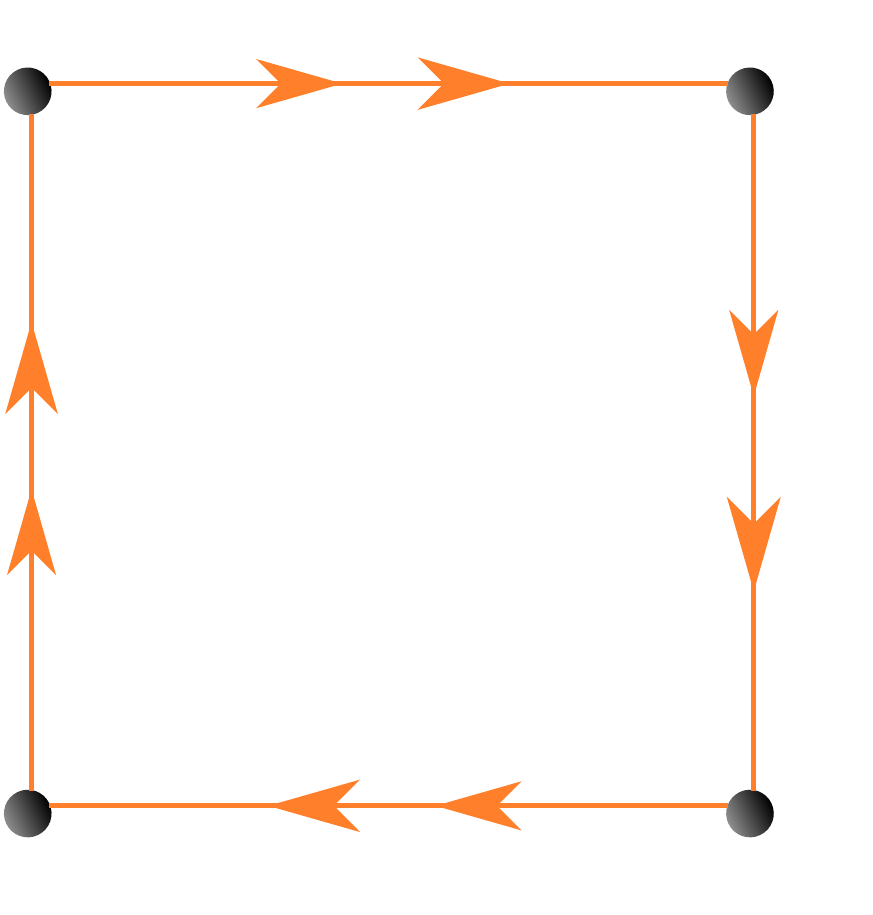}}%
    \put(-0.00475524,1.00466543){\color[rgb]{0,0,0}\makebox(0,0)[lb]{\smash{$k_1=k$}}}%
    \put(0.75225074,1.00466541){\color[rgb]{0,0,0}\makebox(0,0)[lb]{\smash{$k_2=k$}}}%
    \put(-0.00475526,0.01351053){\color[rgb]{0,0,0}\makebox(0,0)[lb]{\smash{$k_4=-k$}}}%
    \put(0.72042466,0.01351053){\color[rgb]{0,0,0}\makebox(0,0)[lb]{\smash{$k_3=-k$}}}%
  \end{picture}%
\endgroup

  \caption {The Chern-Simons quiver gauge theory dual to $AdS_4 \times Q^{2,2,2}/\Z_k$ as proposed in \cite{Franco:2009sp,Davey:2009sr}.\label {Q222Figure}}
\end {figure}
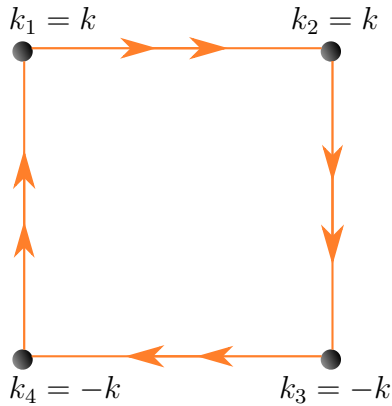
Let us also note that this volume is the same as that of a $\Z_k$ orbifold of the Sasaki-Einstein space $Q^{2, 2, 2}$, which in turn is a $\Z_2$ orbifold of the coset space $SU(2) \times SU(2) \times SU(2) / \left( U(1) \times U(1) \right)$.  If we denote the generators of the three $SU(2)$ factors by $\vec{J}_A$, $\vec{J}_B$, and $\vec{J}_C$, then the two $U(1)$ groups we are modding out by are generated by $J_{A3} + J_{B3}$ and $J_{A3} + J_{C3}$.  $Q^{2, 2, 2}$ admits a toric Sasaki-Einstein metric, and a proposal for the Chern-Simons quiver gauge theory dual to $AdS_4 \times Q^{2, 2, 2}/\Z_k$ was made in \cite{Franco:2009sp,Davey:2009sr}.  This proposal is quite similar to the $(k, k, -k, -k)$ nonchiral quiver in figure~\ref{SquareQuiver}, except it is chiral---see figure~\ref{Q222Figure}. Because of the chiral nature of the quiver, the corresponding matrix model that follows from \cite{Kapustin:2009kz} is somewhat different. Its analysis is beyond the scope of this paper.

\subsection{Extremization of the Free Energy Functional and Symmetries}
\label{SYMMETRIES}

Since the free energy functional \eqref{FNecklace} depends only on differences between consecutive $y_a$, we find it convenient to introduce the notation $\delta y_a = y_{a-1} - y_a$ and to write $k_a = q_{a+1}-q_a$ as in eq.~\eqref{kToq}.  Equation~\eqref{FNecklace} becomes
 \es{FNecklaceAgain}{
  F[\rho, \delta y_a] = \frac{N^{3/2}}{2\pi}  \int dx\, &x \rho(x) \sum_{a=1}^p
     q_a  \delta y_a(x)
     + \frac{N^{3/2}}2 \int dx\, \rho(x)^2 \sum_{a=1}^p f( \delta y_a(x))
     + o(N^{3/2})  \,.
 }
This expression should be extremized over the set
 \es{extremizationSet}{
  {\cal C} =  \left\{ (\rho, \delta y_a): \int dx\, \rho(x) = 1; \rho(x) \geq 0 \text{ and } \sum_{a=1}^p \delta y_a(x) = 0 \text{ pointwise} \right\} \,.
 }

Since $\sum_{a = 1}^p \delta y_a = 0$, one could either use this constraint to solve for one of the $\delta y_a$ and extremize \eqref{FNecklaceAgain} only with respect to the remaining ones, or, as we will do, one could introduce a Lagrange multiplier $\nu(x)$ that enforces the constraint and treat all $\delta y_a$ on equal footing.  Because of the normalization constraint \eqref{rhoNormalization} we also need a Lagrange multiplier $\mu$.  We therefore will extremize
 \es{FWithMultipliers}{
    \tilde F[\rho, \delta y_a] = F[\rho, \delta y_a] - \frac{N^{3/2}}{2 \pi} \mu \left( \int dx\, \rho(x) - 1 \right)
     - \frac{N^{3/2}}{2 \pi} \int dx\, \rho(x) \nu(x) \sum_{a = 1}^p \delta y_a(x)
 }
instead of \eqref{FNecklaceAgain}.  Suppose a saddle point exists.  As long as $\rho(x)>0$, the saddle-point eigenvalue distribution should satisfy the equations
\begin {subequations}
\label{rhozaeqs}
\begin{align}
\sum_{a=1}^p  \left[ 2 \pi f(\delta y_a(x)) \rho(x) + \left(q_a x-\nu(x)\right) \, \delta y_a(x)  \right] &= \mu \,,
\label{rhoeq} \\
\pi f'(\delta y_a(x)) \rho(x)  + q_a x &= \nu(x) \,.
\label{zaeq}
\end{align}
\end {subequations}

The extremization problem has the following discrete symmetries:
 \begin{itemize}
  \item The free energy functional \eqref{FNecklaceAgain} has a $\Z_2$ symmetry under which $q_a$ and $\delta y_a$ all change sign, so in the large $N$ limit the partition function and the free energy are also invariant under sending $q_a \to -q_a$ for all $a$.  This symmetry acts as $k_a \to -k_a$ and is therefore a parity transformation.
  \item Equation~\eqref{FNecklaceAgain} is invariant under an overall shift of all the $q_a$.  This symmetry was to be expected given that, after all, the original integral \eqref{NecklaceModel} depends only on $k_a$, which are differences of consecutive $q_a$.
  \item Interestingly, the free energy functional we are extremizing is invariant under permutations of the $q_a$ and $\delta y_a$, so the partition function and the free energy will also be invariant under permutations of the $q_a$.  Up to order $O(N^0)$ shifts in the ranks of the gauge groups, which should be dropped in the large $N$ limit we are taking, such permutations correspond to Seiberg dualities in the ${\cal N}=2$ Chern-Simons gauge theories \cite{Aharony:2008gk,Giveon:2008zn, Amariti:2009rb}.
 \end{itemize}

Some of the symmetries discussed above correspond to the action of the dihedral group $D_p$ on the CS levels $k_a$.  Our formalism shows that to leading order in $N$ the free energy is in fact invariant under a larger symmetry group that acts on the $q_a$ and that includes the dihedral group.

\subsection{Three-Node Quivers}

Let us now compute the free energy for arbitrary three-node quivers with CS levels $(k_1, k_2, k_3)$ satisfying $k_1 + k_2 + k_3 = 0$.  Since the $k_a$ sum to zero, two of them must have the same sign and be smaller in absolute value than the third.  Let us begin by studying the particular case where $k_2 \geq k_1 \geq 0$ and $k_3 < 0$.  For simplicity, we choose $\sum_{a = 1}^3 q_a = 0$, which implies
 \es{qTok}{
  q_1 = -\frac{2 k_1 + k_2}{3} \,,\qquad
   q_2 = \frac{k_1 - k_2}{3} \,, \qquad
   q_3 = \frac{k_1 + 2 k_2}{3} \,,
 }
and we have $q_3 > 0 > q_2 \geq q_1$ and $\abs{q_3} > \abs{q_1} \geq \abs{q_2}$.  The solution to eqs.~\eqref{rhozaeqs} is symmetric about $x=\delta y_a=0$, and when $x\geq 0$ it breaks into three regions:
\begin {subequations}
\begin{align}
0 \le x \le \frac{\mu}{3 \pi q_3} &: \\
\delta y_a &= \frac{3 \pi^2 x q_a}{\mu} \,, \qquad
\rho = \frac{\mu}{6 \pi^3} \,,\nonumber\\
\frac{\mu}{3 \pi q_3} \le x \le -\frac{\mu}{3 \pi  q_1} &:\\
\delta y_1 &= \frac{(q_1 - q_2)x}{4 \pi \rho} - \frac{\pi}{2}  \,, \qquad
\delta y_2 = \frac{(q_2 - q_1)x}{4 \pi \rho} - \frac{\pi}{2} \,, \qquad
\delta y_3 = \pi \,,  \nonumber \\
\rho &= \frac{2 \mu - 3 \pi q_3 x}{6\pi^3} \,, \nonumber \\
\end {align}
\begin {align}
- \frac{\mu}{3 \pi q_1} \le x \le \frac{\mu}{\pi(q_3 - q_1)} &: \\
\delta y_1 &= -\pi \,,\qquad
\delta y_2 = 0 \,,\qquad
\delta y_3 = \pi  \,,\nonumber \\
\rho &= \frac{\mu+(q_1 - q_3)\pi x }{2 \pi^3} \,.\nonumber
\end{align}
\end {subequations}
\begin {figure} [tbh]
  \center\includegraphics[width=\textwidth] {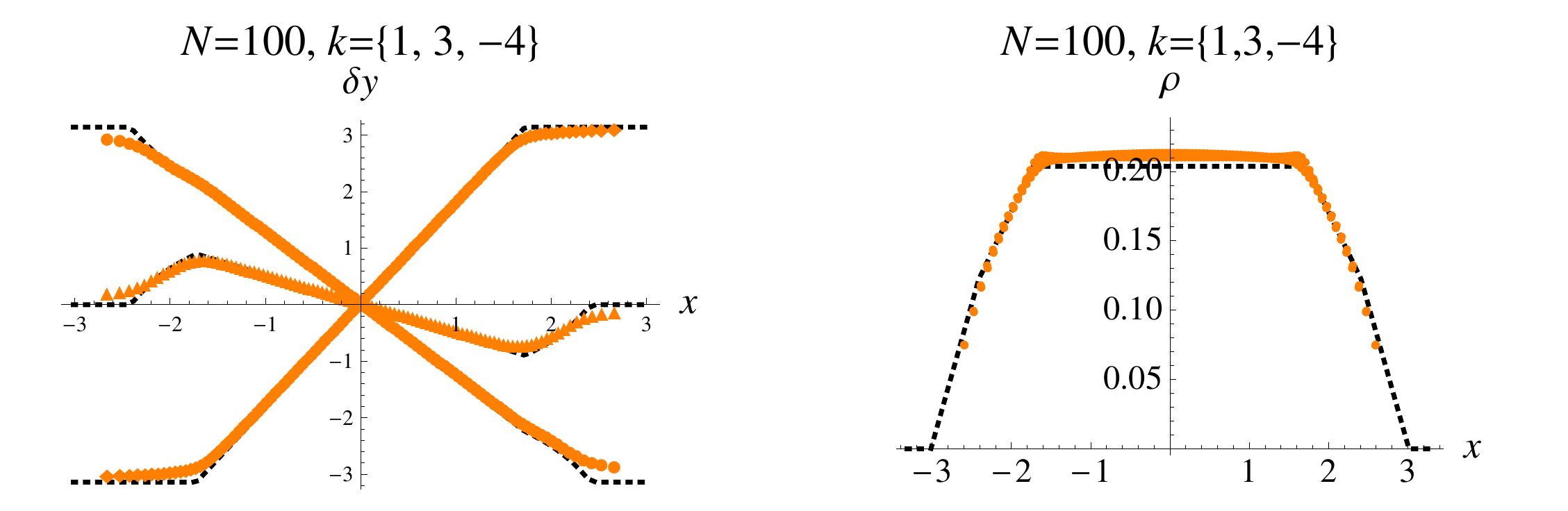}
  \caption {Comparison between numerics and analytical results for a three-node quiver.  The dotted black lines represent the analytical large $N$ approximation, while the orange dots represent numerical results.
  \label {ThreeNodePlots}}
\end {figure}
The first region ends when one of the three differences $\delta y_a$ reaches $\pm \pi$.  The relations between the $q_a$ imply that at the end of the first region $\delta y_3 = \pi$, while $\abs{\delta y_1} = \pi \abs{q_1} / \abs{q_3} < \pi$ and $\abs{\delta y_2} = \pi \abs{q_2} / \abs{q_3} < \pi$.  Throughout the second region $\delta y_3 = \pi$.  The second region ends when $\delta y_1$ or $\delta y_2$ reaches $\pm \pi$.  When $q_1 = q_2$, the third region is absent.  When $q_1<q_2<0$, in this region $\delta y_2$ is monotonically increasing and $\delta y_1$ is monotonically decreasing, and since $\sum \delta y_a = 0$ it must be that $\delta y_1$ reaches $-\pi$ next.  In the third region the $\delta y_a$ are all constant and the density $\rho$ decreases linearly to zero.  See figure~\ref{ThreeNodePlots} for a particular example.

The normalization condition on $\rho$ yields
\be
\mu = \pi^2 \sqrt{\frac{ 18 q_1 (q_1 + q_2) (2 q_1 + q_2)}{q_2^2-5 q_1^2 - 5 q_1 q_2 }}
  = \pi^2 \sqrt{ \frac{2 (k_1 + k_2) (k_2 - k_3) (k_1 - k_3)}
  {(k_1 k_2 - k_1 k_3 - k_2 k_3)} } \,.
\ee
Performing the integral \eqref{FNecklace}, one obtains
 \es{GotFThree}{
   F = \frac{N^{3/2} \mu }{3\pi}
    =  \frac{N^{3/2} \pi \sqrt{2}}{3} \sqrt{\frac{
    (k_1 + k_2) (k_2 - k_3) (k_1 - k_3)}
    {k_1 k_2 - k_1 k_3 - k_2 k_3}}
 \,.
 }

Given the free energy in the case $k_3 < 0 < k_1 \leq k_2$, it is actually possible to compute the free energy for any three-node quivers.  Indeed, since in the case where there are only three nodes a permutation of the $k_a$ can be thought of as a relabeling of the nodes, the free energy must be invariant under all such permutations.  In addition, the free energy must be invariant under sending $k_a \to -k_a$ according to the second discrete symmetry discussed at the end of section~\ref{SYMMETRIES}.  Combining these two properties, one can find the free energy of an arbitrary quiver with CS levels $k_a$ by constructing the new CS levels $\tilde k_1 = \min (\abs{k_1}, \abs{k_2}, \abs{k_3})$,  $\tilde k_3 = -\max (\abs{k_1}, \abs{k_2}, \abs{k_3})$, and $\tilde k_2 = -\tilde k_1 - \tilde k_3$ that satisfy $\tilde k_3 < 0 < \tilde k_1 \leq \tilde k_2$ and for which eq.~\eqref{GotFThree} holds.  The unique extension of \eqref{GotFThree} that gives the correct answer for arbitrary CS levels is
 \es{FThreeGeneral}{
F =  \frac{N^{3/2} \pi \sqrt{2}}{3} \sqrt{\frac{
    (\abs{k_1} + \abs{k_2}) (\abs{k_2} + \abs{k_3}) (\abs{k_1} + \abs{k_3})}
    {\abs{k_1} \abs{k_2} + \abs{k_1} \abs{k_3} + \abs{k_2}  \abs{k_3}}} \,.
 }

Quite remarkably, this formula, whose derivation is based solely on gauge theory arguments, agrees with the supergravity prediction:  Using \eqref{FExpectation}, one can reproduce the volume of a $\Z_{\gcd \{k_1, k_2, k_3 \}}$ orbifold of a compact Eschenburg space.  The Eschenburg space is specified by three relatively prime integers $t_a$, and its volume is \cite{Lee:2006ys}
 \es{volYThreeLeeYee}{
  \frac{\Vol(S(t_1, t_2, t_3))}{\Vol(S^7)} =
    \frac{t_1 t_2 + t_1 t_3 + t_2 t_3}{
    (t_1 + t_2) (t_2 + t_3) (t_1 + t_3)}  \,.
 }
In terms of the $k_a$, the positive integers $t_a$ are $t_a = \abs{k_a} / \gcd\{k_1, k_2, k_3\}$ \cite{Jafferis:2008qz}, so
 \es{volYThree}{
  \frac{\Vol(Y)}{\Vol(S^7)} = \frac{1}{\gcd\{k_1, k_2, k_3 \}} \frac{\Vol(S(t_1, t_2, t_3))}{\Vol(S^7)} = 
    \frac{\abs{k_1} \abs{k_2} + \abs{k_1} \abs{k_3} + \abs{k_2}  \abs{k_3}}{
    (\abs{k_1} + \abs{k_2}) (\abs{k_2} + \abs{k_3}) (\abs{k_1} + \abs{k_3})}  \,,
 }
in agreement with \eqref{FExpectation} and \eqref{FThreeGeneral}.

\subsection{General Four-Node Quivers}

We can also compute the leading large $N$ contribution to the free energy for arbitrary four-node quivers.  Let us first examine the case where $q_4 \geq q_2 \geq q_1 \geq q_3$ and $\abs{q_4}$ is the largest among the $q_a$.   It is convenient to require $\sum_{a=1}^4 q_a = 0$ since many of the intermediate formulae simplify under this assumption.  Then we have $q_4 > 0 \geq q_1 \geq q_3$ and $\abs{q_4} \geq \abs{q_3} \geq \abs{q_1} \geq \abs{q_2}$.  As in the three-node case, the solution to eqs.~\eqref{rhozaeqs} is symmetric about $x=\delta y_a = 0$, and when $x\geq 0$ it breaks into three regions:
\begin {subequations}
\begin{align}
0 \le x \le \frac{\mu}{4 \pi q_4} &: \\
\delta y_a &= \frac{4 \pi^2 x q_a }{\mu} \,,\qquad
\rho = \frac{\mu}{8 \pi^3} \,, \nonumber \\
\frac{\mu}{4 \pi q_4} \le x \le -\frac{\mu}{4 \pi  q_3} &:\\
\delta y_1 &= \frac{(3q_1 + q_4)x}{6 \pi \rho} - \frac{\pi}{3}  \,,\qquad
\delta y_2 = \frac{(3q_2 + q_4)x}{6 \pi \rho} - \frac{\pi}{3} \,, \nonumber \\
\delta y_3 &= \frac{(3q_3 + q_4)x}{6 \pi \rho} - \frac{\pi}{3} \,,\qquad
\delta y_4 = \pi \,,\qquad \rho = \frac{3 \mu - 4 \pi q_4 x}{16\pi^3}  \,, \nonumber \\
- \frac{\mu}{4 \pi q_3} \le x \le \frac{\mu}{2\pi(q_2 + q_4)} &: \\
\delta y_1 &= \frac{(q_1 - q_2) x}{4 \pi \rho} \,,\qquad
\delta y_2 = \frac{(q_2 - q_1)x}{4 \pi \rho} \,, \nonumber \\
\delta y_3 &= -\pi  \,,\qquad
\delta y_4 = \pi \,,\qquad
\rho = \frac{\mu+(q_3 - q_4)\pi x }{4 \pi^3} \,.\nonumber
\end{align}
\end {subequations}
\begin {figure} [tbh]
  \center\includegraphics[width=\textwidth] {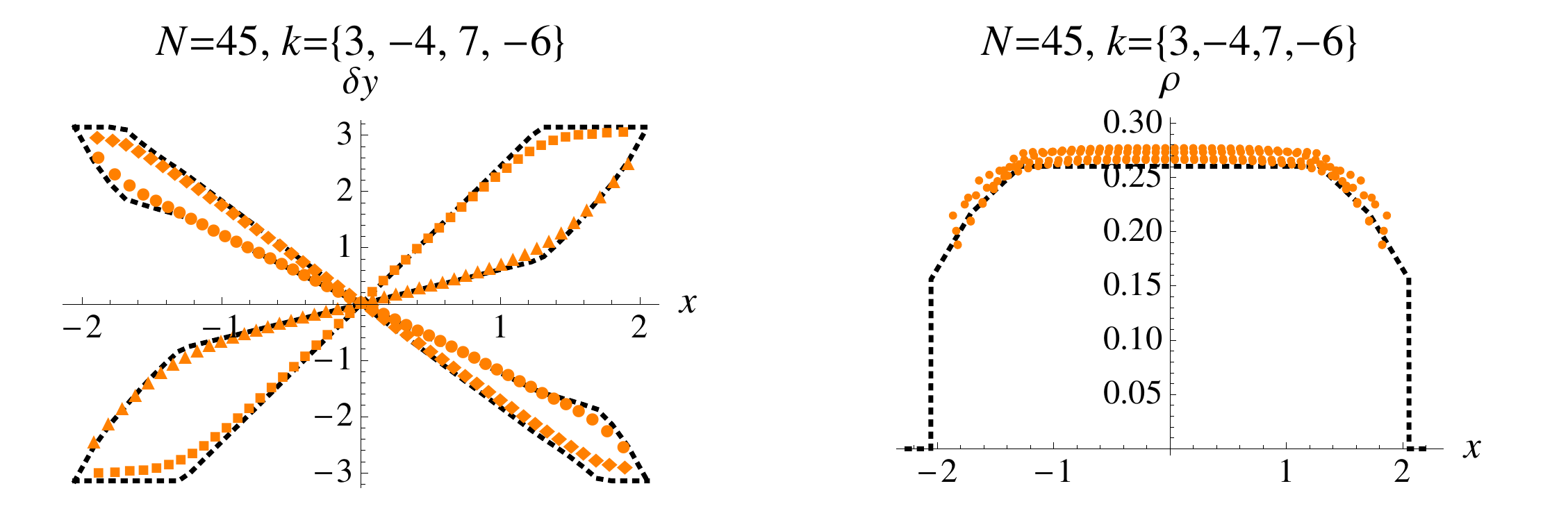}
  \caption {Comparison between numerics and analytical results for a four-node quiver.  The dotted black lines represent the analytical large $N$ approximation, while the orange dots represent numerical results.
  \label {FourNodePlots}}
\end {figure}%
The first region ends where $\delta y_4$ reaches $\pi$.  At this endpoint $\abs{\delta y_a} = \pi \abs{q_a} / \abs{q_4} \leq \pi$ for $a = 1, 2, 3$.  The second region ends where $\delta y_3 = -\pi$.  At this endpoint $\delta y_1 =\pi (q_1 - q_2) / (q_1 + q_2 - 2 q_3)$, and since $q_1 \geq q_3$ and $q_2  \geq q_3$, by the triangle inequality it follows that $\abs{q_1 - q_2} \leq q_1 + q_2 - 2 q_3$, so $\abs{\delta y_1} \leq \pi$.  Similarly, $\abs{\delta y_2}\leq \pi$ also.  Lastly, if $q_2 = q_1$, the third region does not exist. When $q_2>q_1$ and $q_1 < 0$, $\delta y_1$ is monotonically decreasing and $\delta y_2$ is monotonically increasing in the third region, and this region ends where $\delta y_1 = -\pi$ and $\delta y_2 = \pi$. See figure~\ref {FourNodePlots} for an example.

From $\int dx\, \rho(x) = 1$, one can find that $\mu$ is given by
 \es{muFourNode}{
  \frac{8 \pi^2}{\mu} = \sqrt{
   \frac{1}{q_3} - \frac{1}{q_4} + \frac{4(q_2 + q_3)}{(q_2 + q_4)^2} + \frac{12}{q_2 + q_4} } \,.
 }
The free energy is
 \es{GotFreeFour}{
  F = \frac{N^{3/2}  \mu}{3\pi}
   =  \frac{8\pi N^{3/2} }{3} \left( \frac{1}{q_3} - \frac{1}{q_4} + \frac{4(q_2 + q_3)}{(q_2 + q_4)^2} + \frac{12}{q_2 + q_4} \right)^{-1/2} \,.
 }

Given eq.~\eqref{GotFreeFour}, one can use the symmetries we discussed at the end of section~\ref{SYMMETRIES} to compute the free energy of a quiver gauge theory with arbitrary $q_a$.  Indeed, one can define $\tilde q_a$ to be a permutation of the four numbers $q_a - \frac 14 \sum_{b = 1}^4 q_b$ that gives $\abs{\tilde q_4} \geq \abs{\tilde q_3} \geq \abs{\tilde q_1} \geq \abs{\tilde q_2}$.  If $\tilde q_4$ is negative, one should flip the sign of all $\tilde q_a$, so we can assume $\tilde q_4>0$.  By construction, the $\tilde q_a$ sum to zero, so the second and third largest in absolute value, namely $\tilde q_3$ and $\tilde q_1$, are negative.  Therefore, the $\tilde q_a$ satisfy all the assumptions under which eq.~\eqref{GotFreeFour} was derived, and since the free energy does not change in going from $q_a$ to $\tilde q_a$, one can plug the $\tilde q_a$ into eq.~\eqref{GotFreeFour} to find the free energy of an arbitrary four-node quiver theory.  The unique extension of \eqref{GotFreeFour} to arbitrary $q_a$ can also be written as
 \es{ExtensionFour}{
   F = \frac{N^{3/2} \pi \sqrt{2}}{3} \sqrt{
    \frac{\prod_{a = 1}^4 \left( \sum_{b = 1}^4 \abs{q_{ab}} \right) }
    {\sum_{(a, b) \neq (c, d) \neq (e, f)} \abs{q_{ab}} \abs{q_{cd}} \abs{ q_{ef} }
     - \sum_{(a, b, c)} \abs{ q_{ab}} \abs{ q_{bc}} \abs{ q_{ca}}  } } \,,
 }
where $q_{ab}$ denotes $q_a - q_b$, and in the denominator the first sum is over distinct unordered pairs of numbers from $1$ to $4$ while the second sum is over unordered triplets.  Using eq.~\eqref{FExpectation}, we obtain a prediction for the volume of the compact space $Y$:
 \es{VolPredictionFour}{
  \frac{\Vol(Y)}{\Vol(S^7)}
   =  \frac
    {\sum_{(a, b) \neq (c, d) \neq (e, f)} \abs{q_{ab}} \abs{q_{cd}} \abs{ q_{ef} }
     - \sum_{(a, b, c)} \abs{ q_{ab}} \abs{ q_{bc}} \abs{ q_{ca}}  }
     {\prod_{a = 1}^4 \left( \sum_{b = 1}^4 \abs{q_{ab}} \right) } \,.
 }

\section{A General Formula and its Tests}

Equations~\eqref{ExtensionFour} and \eqref{VolPredictionFour} suggest a
generalization to arbitrary $p$-node quivers.
Note first that the numerator of eq.~\eqref{VolPredictionFour} is a sum over all possible graphs with 4 nodes and 3 edges from which we subtract the sum over all cyclic graphs with 4 nodes and 3 edges, yielding a sum over all possible trees.

We conjecture that for a $p$-node quiver,
the volume of the tri-Sasaki Einstein space $Y$ (normalized so that $R_{mn} = 6 g_{mn}$) is given by
 \es{ConjectureVolume}{
  \frac{\Vol(Y)}{\Vol(S^7)} =
  \frac{ \sum_{(V,E) \in {\cal T}} \prod_{(a,b) \in E} |q_a - q_b|}
  {\prod_{a = 1}^p \Bigl[ \sum_{b = 1}^p \abs{q_a - q_b} \Bigr] } \,,
 }
where $\mathcal T$ is the set of all trees (acyclic connected graphs)
with nodes $V = \{1, 2, \ldots, p\}$ and edges
 \es{Edges}{
  E = \{(a_1, b_1), (a_2, b_2), \ldots, (a_{p-1}, b_{p-1})\} \,.
 }
A standard result in graph theory states that trees with $p$ nodes have $p-1$ edges.

The conjecture in eq.~\eqref{ConjectureVolume} is consistent with the results from two-, three-, and four-node quivers, and we also checked it for five- and six-node quivers.  This formula is invariant under all the symmetries discussed at the end of section~\ref{SYMMETRIES}.  In particular, a quite nontrivial check of our approach is that this formula is invariant under the Seiberg dualities described in \cite{Aharony:2008gk,Giveon:2008zn, Amariti:2009rb}.  The connection we observe between large $N$ matrix integrals and sums over the tree graphs is reminiscent of the connection between matrix models for 2-d quantum gravity and the Kontsevich matrix model which generates ribbon graphs \cite{Kontsevich:1992}.

An integral representation of volumes of tri-Sasaki Einstein spaces was given by Yee \cite{Yee:2006ba}.  In general, our spaces $Y$ are $\Z_k$ orbifolds of those considered in \cite{Yee:2006ba}, where $k = \gcd\{k_a\}$.  To simplify the following discussion, let us focus on the $k=1$ case.  In this case \cite{Yee:2006ba},
\be
\Vol(Y) = \frac{2^{p-2} \pi^4}{3 \Vol\left(U(1)^{p-2}\right)} \int \prod_{j=1}^{p-2} d \phi^j \prod_{a=1}^{p} \frac{1}{1 + \left( \sum_{j=1}^{p-2} Q^j_a \phi^j \right)^2} \,.
\label{Yee}
\ee
Here, $\Vol \left(U(1)^{p-2}\right)$ is the volume of a unit cell in the $(p-2)$-dimensional lattice defined by the identifications $\xi_j \sim \xi_j + \eta_j$, where $\eta_j$ satisfy $\sum_{j=1}^{p-2} Q_a^j \eta_j \in 2\pi\mathbb Z$ for all $a = 1,\dotsc,p$.  The $Q^j_a$ span the kernel of
\be
\begin{pmatrix}
1 & 1 & 1 & \cdots & 1 \\
q_1 & q_2 & q_3 & \cdots & q_p
\end{pmatrix}
\,.
\ee
(The $Q^j_a$ are taken to be relatively prime here.)  In the $U(1)^p$ Chern-Simons gauge theory, the $Q^j_a$ are the charges of the bifundamental fields under the unbroken $U(1)^{p-2}$ symmetry \cite{Jafferis:2008qz}.  We can take a spanning set of $\vec Q^j$
to be, for a fixed $j$, $Q^j_1 = q_2 - q_j$, $Q^j_2 = q_j - q_1$, and $Q^j_j = q_1 - q_2$ with all other $Q^j_a = 0$.  For this choice of $Q^j_a$, the volume of $U(1)^{p-2}$ is 
 \es{VolU1}{
  \Vol\left(U(1)^{p-2}\right) = \frac{(2 \pi)^{p-2}}{\abs{q_1 - q_2}^{p-3}} \,.
 }
Note that $\Vol(Y)$ is invariant under permutation of the $q_a$.  Although we have not carried out the integral in general, we can investigate specific cases with ease.  For example, for the choice $\vec q = (3,2,1,2)$, corresponding to the $\vec k = (1,1,-1,-1)$ quiver, both our formulae~\eqref{ConjectureVolume} and~\eqref{Yee} give $\Vol(Y) = \pi^4 / 16$. A more nontrivial choice is $\vec q = (3,2,1,5)$ for which both formulae yield $139 \pi^4 / 4725$.  By evaluating \eqref{Yee} numerically, we were able to check agreement with \eqref{ConjectureVolume} in a number of randomly selected cases for $p = 4, 5$, and $6$.

The volume formula \eqref{ConjectureVolume} is invariant under a shift $q_a\rightarrow q_a+1$. In the type IIB brane construction, which involves a sequence of $(1,q_a)$ 5-branes, this symmetry corresponds to the T transformation of the $SL(2,\Z)$ S-duality group. We could use the $SL(2, \Z)$ symmetry to generalize the free energy to theories whose brane constructions involve general $u_a=(p_a,q_a)$ 5-branes.  This generalization is accomplished by  replacing the differences $\abs{q_a - q_b}$ in the volume formula with $\abs{u_a \wedge u_b} = \abs{p_a q_b - p_b q_a}$. For special cases where some of the $p_a$ vanish, this formula describes theories with fields in the fundamental representation.  For example, for the ABJM model with $N_f$ flavors, corresponding to $u_1=(1,k), u_2=(1,0), u_3=(0, N_f)$, our formula predicts 
 \es{ABJMFlavors}{
  \frac{\Vol(Y)}{\Vol(S^7)} = \frac{2 k + N_f} {2 (k + N_f)^2} \,.
 }
This equation agrees with the explicit matrix model calculation \cite{Santamaria:2010dm} and with the volumes of Eschenburg spaces $S(N_f,N_f, k)$ \cite{Lee:2006ys}.

\section{Discussion}

In this paper we have studied $p$-matrix models describing certain $U(N)^p$ Chern-Simons quiver gauge theories with ${\cal N} = 3$ supersymmetry.  In the large $N$ limit these theories are dual to eleven-dimensional supergravity on $AdS_4\times Y$, where $Y$ is a tri-Sasaki Einstein space.  By finding an analytical large $N$ limit of the matrix integrals, we were able to check the supergravity prediction that the logarithm of the partition function of the gauge theories on $S^3$ should grow as $N^{3/2}$.  In $AdS_4\times Y$ the coefficient of proportionality depends on the volume of the compact spaces $Y$, so we could compare our gauge theory results with the volumes computed earlier using geometric techniques \cite{Lee:2006ys,Yee:2006ba}. These successful comparisons constitute new detailed tests of the $AdS_4$/CFT$_3$ dualities. In eq.~\eqref{ConjectureVolume} we conjectured an explicit combinatorial volume formula for arbitrary $p$.  It should be possible to derive this formula in an independent way using algebraic geometry techniques similar to those in~\cite{Bergman:2001qi}.

Quite generally, the main difficulty in solving matrix models is that the interactions between the eigenvalues are long-ranged, and the saddle-point approximation yields integral equations in the continuum limit.  Remarkably, in solving the models described in this paper, one can set up an approximation scheme where the eigenvalue distributions can be found by solving algebraic equations.  The limit in which the saddle-point equations simplify is the limit of ``large cuts'' where the eigenvalues grow as an appropriate positive power of $N$.  Perhaps the key insight in solving these matrix models was that the long-range forces between the eigenvalues can be made to vanish by choosing the distribution of the real parts of the eigenvalues to be the same for each set of eigenvalues.  The remaining interaction forces between the eigenvalues are short-ranged, and that is the reason why in the right variables the saddle-point equations were local and algebraic in the large $N$ limit.

While we worked in the limit where $N$ is sent to infinity and the Chern-Simons levels $k_a$ are kept fixed, it is of obvious further interest to relax these assumptions and study $1/N$ corrections.  In doing so, a subtle issue that needs a better understanding is the imaginary part of the free energy.  At first sight, the imaginary part in the ABJM model is of order $O(N)$.  On the other hand, one could argue that this imaginary part is only defined modulo $2 \pi$ because a shift of the free energy by an integer multiple of $2 \pi i$ leaves the partition function unchanged.

Another interesting generalization of our results is to solve the matrix model in the scaling limit where the Chern-Simons levels are sent to infinity, with $N/k_a$ kept finite.  One could calculate the free energy as a function of the 't Hooft-like couplings $N/k_a$ and check that, as predicted by the AdS/CFT correspondence, it should interpolate between an $N^2$ behavior at small $N/k_a$ dictated by perturbation theory and the $ k^{1/2} N^{3/2}$ behavior at large $N/k_a$ that we found.   For $p=2$ this check was performed in  \cite{Drukker:2010nc} by computing the resolvent of the matrix model using the techniques developed in \cite{Halmagyi:2003ze}.  We believe a similar check should also be possible for the ${\cal N}=3$ theories studied in this paper, using perhaps similar techniques.  Such an approach should also provide access to the ABJ-like cases where the ranks of the $p$ gauge groups are not equal.

Finally, it would be interesting to investigate whether the large $N$ matrix integrals we have calculated play a role in four-dimensional gauge theories, for example, in the 4-d ``parent theories'' \cite{Klebanov:1998hh,Gubser:1998ia} of the 3-d Chern-Simons models we have studied.

\section*{Acknowledgments}
We thank N.~Halmagyi, N.~Kamburov, J.~Maldacena, M.~Marino, N.~Nekrasov, V.~Pestun, and especially M.~Yamazaki for useful discussions.  This work was supported in part by the US NSF  under Grants No.~PHY-0756966 and No.~PHY-0844827.  The work of CPH was also supported in part by the Sloan Foundation, and that of SSP by Princeton University through a Porter Ogden Jacobus Fellowship and the Compton Fund.  CPH and SSP thank the Galileo Galilei Institute for Theoretical Physics for hospitality and the INFN for partial support during the final stages of this work. IRK is grateful to the Aspen Center for Physics and the Erwin Schr\"odinger Institute in Vienna for hospitality.  SSP is also thankful to the Harvard University Physics Department for hospitality while this work was in progress.

\appendix

\section{Taking the Continuum Limit of the Free Energy}
\label{app:free}

\newcommand {\txext} {\text{ext}}
\newcommand {\txint} {\text{int}}
\newcommand {\txconst} {\text{const}}
We start with the matrix integral eq.~(\ref{NecklaceModel}) and write it as
\be
Z = \int e^{-F(\lambda_{a,i})} \prod_{a,i} d \lambda_{a,i} \,,
\ee
where we divide the free energy into the following three pieces: $
F = F_{\txext} + F_{\txint} + F_{\txconst}$.\footnote {In the following, we will use the product formula $\log (xy) = \log x + \log y$, although this is strictly speaking only correct up to integer multiples of $2\pi i$. The extra contributions would not affect the saddle-point equations.}
We have defined $F_{\txext}$ to be the contribution to the free energy from the external potential
\be
F_{\txext} \equiv -\frac{i}{4\pi} \sum_{a,i} k_a \lambda_{a,i}^2 \,.
\ee
The contribution to $F$ from the eigenvalue interactions is
\be
F_{\txint} \equiv - \log \prod_{a=1}^p
\frac{\prod_{i>j}  4 \sh^2 \left(\frac{\lambda_{a,i} - \lambda_{a,j}}{2}\right)^2 }
{\prod_{i,j} 2 \ch \left(\frac{\lambda_{a,i} - \lambda_{a+1,j}}{2}\right) } \,.
\ee
Finally,
there is an overall normalization $F_{\txconst}$ chosen to be consistent with \cite{Drukker:2010nc}:
\be
F_{\txconst} \equiv p \, ( \log N! + N \log 2 \pi) \,.
\ee
It will turn out that $F_{\txconst}$ does not contribute at leading order in our large $N$ expansion.

To take the continuum limit, we begin with the assumptions justified in the text of the paper that the eigenvalue distributions lie along curves symmetric about the origin
in the complex $\lambda$ plane, that we can order the eigenvalues such that
$\Re\lambda_{a,j} = \Re\lambda_{b,j}$ for any $a$ and $b$, and that $\abs{\Re\lambda_{a,j}} \gg \abs{\Im\lambda_{a,j}}$.  More specifically we assume the following variant of (\ref{lambdaLargeN}):
\be
\lambda_{a,j} = N^{\alpha} x_{j} + i y_{a,j} \,,
\label{lambdaLargeNgen}
\ee
where $\alpha >0$.

Taking the continuum limit of $F_{\txext}$ is straightforward.  The leading term in $N$ cancels because $\sum_a k_a = 0$, and we are left with
\be
F_{\txext} = \frac{N^\alpha}{2\pi} \sum_{a,j} k_a x_{a,j} y_{a,j} + O(N) \,.
\ee
Letting the real parts of the eigenvalue distributions extend
from $-x_*$ to $x_*$, we can introduce an eigenvalue density $\rho(x)$ and
approximate the sum over $j$ as an integral over $x$:
\be
F_{\txext} = \frac{N^{1+\alpha}}{2\pi} \sum_a k_a  \int_{-x_*}^{x_*} x y_a(x) \rho(x) \, dx + O(N) \,.
\ee

Taking the continuum limit of $F_{\text{int}}$ is more involved.
We begin by reorganizing the products:
\be
\begin {split}
F_{\text{int}} &= - \log  \prod_{a=1}^p \left[
 \prod_{i>j}
 \left(
\frac{
4 \sh^2 \left( \frac{\lambda_{a,i} - \lambda_{a,j}}{2} \right)
}
{
 2\ch \left( \frac{\lambda_{a,i} - \lambda_{a+1,j}}{2} \right)
 2 \ch \left( \frac{\lambda_{a,i} - \lambda_{a-1,j}}{2} \right)
}
\right)
\frac{
1}
{
\prod_{i} 2 \ch \left( \frac{\lambda_{a,i} - \lambda_{a+1,i}}{2} \right)
}
 \right] \\
 &= - \log
 \prod_{a=1}^p \left[ \prod_{i>j} \left(
\frac{ (1 - e^{-\lambda_{a,i}+\lambda_{a,j}})^2}
{(1 + e^{-\lambda_{a,i} + \lambda_{a+1,j}}) ( 1 +  e^{-\lambda_{a,i} + \lambda_{a-1,j}}) }
\right)
\frac{1}
{\prod_i 2 \ch \left( \frac{\lambda_{a,i} - \lambda_{a+1,i}}{2} \right) }
\right] \,.
\end{split}
\ee
We then convert the logarithm of the product into a sum over logarithms:
\be
\begin {split}
 F_{\text{int}} &= \sum_{a=1}^p \left[
\sum_{i>j} \sum_{n=1}^\infty
\frac{1}{n}
\Bigl[
2 e^{(-\lambda_{a,i} + \lambda_{a,j})n}
-
(-1)^n \left( e^{(-\lambda_{a,i} + \lambda_{a+1,j})n}
+  e^{(-\lambda_{a,i} + \lambda_{a-1,j})n} \right)
\Bigr] \right.
+ \\
&\qquad
\left.
+ \sum_i
 \log \left(2 \ch \frac{\lambda_{a,i} - \lambda_{a+1,i}}{2}
\right)
\right] \,.
\end{split}
\ee
Making use of the assumption (\ref{lambdaLargeNgen}) and taking the continuum limit, the interaction energy reduces to
\be
\begin {split}
F_{\text{int}} &= \sum_{a=1}^p \int_{-x_*}^{x_*} \left[
N  \log \left(2 \cos \frac{y_a(x) - y_{a+1}(x)}{2}\right)
+ \int_{-x_*}^x \sum_{n=1}^\infty  \frac{N^2}{n}
\left[
2 e^{(-\lambda_a(x) + \lambda_a(x'))n}+
\right.
\right.
\\
&
\left.
- (-1)^n \left( e^{(-\lambda_a(x) + \lambda_{a+1}(x'))n}
+ e^{(-\lambda_a(x) + \lambda_{a-1}(x'))n} \right)
\right]\rho(x') \, dx'
\Bigg]
\rho(x) dx \,.
\end{split}
\ee

We now estimate the integral over $x'$ in the above expression for $F_{\rm int}$.
Consider the following related integral:
\be
\begin {split}
I &= \int_{-x_*}^x e^{(-\lambda_{b}(x) + \lambda_{a}(x')) n} \rho(x') dx' \\
&=
\left.
\frac{1}{n} e^{(-\lambda_b(x) + \lambda_{a}(x'))n} \frac{dx'}{d \lambda_{a}} \rho(x') \right|_{-x_*}^x
- \frac{1}{n} \int_{-x_*}^x e^{(-\lambda_b(x) + \lambda_a(x'))n}
\frac{d}{dx'} \left( \frac{dx'}{d \lambda_a} \rho(x') \right) dx'
\\
&=
 \frac{1}{n}  \frac{dx}{d\lambda_a} \rho(x) e^{(-\lambda_b(x) + \lambda_a(x)) n}
  + \ldots
\,.
\end{split}
\label{relint}
\ee
Given~\eqref{lambdaLargeNgen}, the integral in the second line will be suppressed by a factor of $1/N^{\alpha}$ compared with the boundary term.  The boundary contribution from $x_*$ will be
suppressed by an exponential amount because
$\Re\bigl[\lambda_a(-x_*)\bigr] < \Re\bigl[\lambda_b(x)\bigr]$.
 The last line of~\eqref{relint} reduces to
\be
I = N^{-\alpha}
\frac{\rho(x)}{n} e^{in(y_a(x) - y_b(x))}+ O(N^{-2 \alpha}) \,.
\ee

Introducing the notation $\delta y_a = y_{a-1} - y_a$, the interaction energy reduces to
\be
F_{\txint} =
N^{2-\alpha} \sum_{a=1}^p \int\limits_{-x_*}^{x_*}
 \sum_{n=1}^\infty \Bigl[
 2 - (-1)^n \left( e^{-in \, \delta y_{a+1}(x)} + e^{in \, \delta y_a(x)} \right) \Bigr] \frac{\rho(x)^2}{n^2} \, dx
 + O(N^{2-2\alpha}, N) \,.
 \ee
 We are tacitly assuming that $\alpha < 1$ and so can drop the order $N$ term from the energy.
 Reorganizing the sum over $a$, we can write this energy as
 \be
 F_{\txint} =
\frac{N^{2-\alpha}}{2} \sum_{a=1}^p \int_{-x_*}^{x_*}
 f(\delta y_a) \rho(x)^2 \, dx
 + O(N^{2-2\alpha}, N) \,.
 \ee
where we have defined the function
\be
f(y) \equiv
 \sum_{n=1}^\infty \frac{4}{n^2}
\bigl[1 - (-1)^n \cos ny
\bigr]  \,.
\ee
Clearly $f$ is a periodic function of $y$ with period $2\pi$.
Recall the Fourier series expansion for $y^2$ in the domain $-\pi < y < \pi$:
\be
y^2 = \sum_{n=1}^\infty \frac{4 \,(-1)^n}{n^2} \cos n y + 2 \zeta(2) \,.
\ee
In the fundamental domain $-\pi < y < \pi$, the function $f$ is thus
\be
f(y) =\pi^2- y^2 \,.
\ee

\section{A More Detailed Check for ABJM Theory}
\label{MOREDETAILS}

We explore solutions to eqs.~\eqref{eoms} where $|y| \geq \pi/2$.  Based on the numerical results in section~\ref{NUMERICS}, we expect the eigenvalue distributions to be invariant under $\lambda_i \to -\lambda_i$ and $\tilde \lambda_i \to -\tilde \lambda_i$, which implies $y(-x) = -y(x)$ and $\rho(-x) = \rho(x)$.   Assuming $k>0$, we will focus only on the $x\geq 0$ region. Plugging \eqref{fABJM} into \eqref{eoms} one obtains
 \es{rhobRegion1}{
   \rho(x) = {\mu \over 4 \pi^3} \,, \qquad y(x) = {\pi^2 k x \over 2 \mu}\,, \qquad
   \text{if} \quad \abs {y(x)} \leq \frac \pi2 \,,
 }
and
 \es{rhobRegion2}{
   \rho(x) = {k (\mu - 2 k x \pi) \over 4 \pi^3} \,, \qquad y(x) = {\pi (2 \mu - 3 k x \pi) \over  2\mu - 4 k x \pi}\,, \qquad
   \text{if} \quad {\pi \over 2} \leq y(x) \leq {3 \pi \over 2} \,,
 }
and so on.   From eq.~\eqref{rhobRegion1} we infer that $\mu>0$ and $y(x)\geq 0$ if $x\geq 0$.  We could have in principle also allowed $y(x) = \pi/2$ over some range of $x$, but then the first equation in \eqref{eoms} would imply that $x =  \mu / \pi k$, so $y(x)$ could equal $\pi/2$ only on a set of measure zero.

Assuming a connected distribution of eigenvalues of each type where $\rho$ is supported on $[-x_{*}, x_{*}]$ for some $x_{*}>0$, there are two possibilities:  Either $y(x_{*}) > \pi/2$ or $y(x_{*}) \leq \pi/2$.  Assuming $y(x_{*}) > \pi/2$ we immediately reach a contradiction.  Indeed,
consider the point $x_{\pi/2} = \mu/\pi k$ where $y(x_{\pi/2}) = \pi/2$ and
eq.~\eqref{rhobRegion1} joins onto eq.~\eqref{rhobRegion2}.
For $x > x_{\pi/2}$, eq.~\eqref{rhobRegion2} implies $\rho(x) < 0$,
which contradicts the assumption that $\rho(x)>0$.  It must be that $\rho(x) = 0$ for $x > x_{\pi/2}$ and thus $\abs{y(x)} \leq \pi/2$ for our eigenvalue distribution.

\bibliographystyle{ssg}
\bibliography{matrix}

\begingroup\raggedright\begin{thebibliography}{10}

\bibitem{Maldacena:1997re}
J.~M. Maldacena, ``{The large $N$ limit of superconformal field theories and
  supergravity},'' {\em Adv. Theor. Math. Phys.} {\bf 2} (1998) 231--252,
  \href{http://xxx.lanl.gov/abs/hep-th/9711200}{{\tt hep-th/9711200}}.

\bibitem{Gubser:1998bc}
S.~S. Gubser, I.~R. Klebanov, and A.~M. Polyakov, ``Gauge theory correlators
  from non-critical string theory,'' {\em Phys. Lett.} {\bf B428} (1998)
  105--114, \href{http://xxx.lanl.gov/abs/hep-th/9802109}{{\tt
  hep-th/9802109}}.

\bibitem{Witten:1998qj}
E.~Witten, ``{Anti-de Sitter space and holography},'' {\em Adv. Theor. Math.
  Phys.} {\bf 2} (1998) 253--291,
  \href{http://xxx.lanl.gov/abs/hep-th/9802150}{{\tt hep-th/9802150}}.

\bibitem{Klebanov:1996un}
I.~R. Klebanov and A.~A. Tseytlin, ``{Entropy of near extremal black
  $p$-branes},'' {\em Nucl.Phys.} {\bf B475} (1996) 164--178,
  \href{http://xxx.lanl.gov/abs/hep-th/9604089}{{\tt hep-th/9604089}}.

\bibitem{Drukker:2010nc}
N.~Drukker, M.~Marino, and P.~Putrov, ``{From weak to strong coupling in ABJM
  theory},'' \href{http://xxx.lanl.gov/abs/1007.3837}{{\tt 1007.3837}}.

\bibitem{Aharony:2008ug}
O.~Aharony, O.~Bergman, D.~L. Jafferis, and J.~Maldacena, ``{${\cal N}=6$
  superconformal Chern-Simons-matter theories, M2-branes and their gravity
  duals},'' {\em JHEP} {\bf 10} (2008) 091,
  \href{http://xxx.lanl.gov/abs/0806.1218}{{\tt 0806.1218}}.

\bibitem{Kapustin:2009kz}
A.~Kapustin, B.~Willett, and I.~Yaakov, ``{Exact Results for Wilson Loops in
  Superconformal Chern-Simons Theories with Matter},'' {\em JHEP} {\bf 1003}
  (2010) 089, \href{http://xxx.lanl.gov/abs/0909.4559}{{\tt 0909.4559}}.

\bibitem{Pestun:2007rz}
V.~Pestun, ``{Localization of gauge theory on a four-sphere and supersymmetric
  Wilson loops},'' \href{http://xxx.lanl.gov/abs/0712.2824}{{\tt 0712.2824}}.

\bibitem{Marino:2009jd}
M.~Marino and P.~Putrov, ``{Exact Results in ABJM Theory from Topological
  Strings},'' {\em JHEP} {\bf 1006} (2010) 011,
  \href{http://xxx.lanl.gov/abs/0912.3074}{{\tt 0912.3074}}.

\bibitem{Halmagyi:2003ze}
N.~Halmagyi and V.~Yasnov, ``{The Spectral curve of the lens space matrix
  model},'' {\em JHEP} {\bf 0911} (2009) 104,
  \href{http://xxx.lanl.gov/abs/hep-th/0311117}{{\tt hep-th/0311117}}.

\bibitem{Suyama:2010hr}
T.~Suyama, ``{On Large $N$ Solution of Gaiotto-Tomasiello Theory},'' {\em JHEP}
  {\bf 1010} (2010) 101, \href{http://xxx.lanl.gov/abs/1008.3950}{{\tt
  1008.3950}}.

\bibitem{Imamura:2008nn}
Y.~Imamura and K.~Kimura, ``{On the moduli space of elliptic
  Maxwell-Chern-Simons theories},'' {\em Prog. Theor. Phys.} {\bf 120} (2008)
  509--523, \href{http://xxx.lanl.gov/abs/0806.3727}{{\tt 0806.3727}}.

\bibitem{Jafferis:2008qz}
D.~L. Jafferis and A.~Tomasiello, ``{A simple class of ${\cal N}=3$
  gauge/gravity duals},'' {\em JHEP} {\bf 10} (2008) 101,
  \href{http://xxx.lanl.gov/abs/0808.0864}{{\tt 0808.0864}}.

\bibitem{Boyer:1993}
C.~P. Boyer, K.~Galicki, and B.~M. Mann, ``{Quaternionic reduction and Einstein
  manifolds},'' {\em Comm. Anal. Geom.} {\bf 1} (1993) 229--279.

\bibitem{Boyer:1994}
C.~P. Boyer, K.~Galicki, and B.~M. Mann, ``{The geometry and topology of
  3-Sasakian manifolds},'' {\em J. Reine Angew. Math} {\bf 455} (1994)
  193--220.

\bibitem{Boyer:1998sf}
C.~P. Boyer and K.~Galicki, ``{3 - Sasakian manifolds},'' {\em Surveys
  Diff.Geom.} {\bf 7} (1999) 123--184,
  \href{http://xxx.lanl.gov/abs/hep-th/9810250}{{\tt hep-th/9810250}}. To
  appear in 'Essays on Einstein Manifolds', M. Wang and C. Lebrun, eds.

\bibitem{Emparan:1999pm}
R.~Emparan, C.~V. Johnson, and R.~C. Myers, ``{Surface terms as counterterms in
  the AdS/CFT correspondence},'' {\em Phys. Rev.} {\bf D60} (1999) 104001,
  \href{http://xxx.lanl.gov/abs/hep-th/9903238}{{\tt hep-th/9903238}}.

\bibitem{Eschenburg}
J.~H. Eschenburg, ``{New examples of manifolds with strictly positive
  curvature},'' {\em Invent. Math.} {\bf 66} (1982) 469--480.

\bibitem{Lee:2006ys}
K.-M. Lee and H.-U. Yee, ``{New $AdS_4\times X_7$ Geometries with ${\cal N}=6$
  in M-theory},'' {\em JHEP} {\bf 03} (2007) 012,
  \href{http://xxx.lanl.gov/abs/hep-th/0605214}{{\tt hep-th/0605214}}.

\bibitem{Yee:2006ba}
H.-U. Yee, ``{AdS/CFT with Tri-Sasakian Manifolds},'' {\em Nucl.Phys.} {\bf
  B774} (2007) 232--255, \href{http://xxx.lanl.gov/abs/hep-th/0612002}{{\tt
  hep-th/0612002}}.

\bibitem{Aharony:2008gk}
O.~Aharony, O.~Bergman, and D.~L. Jafferis, ``{Fractional M2-branes},'' {\em
  JHEP} {\bf 0811} (2008) 043, \href{http://xxx.lanl.gov/abs/0807.4924}{{\tt
  0807.4924}}.

\bibitem{Giveon:2008zn}
A.~Giveon and D.~Kutasov, ``{Seiberg Duality in Chern-Simons Theory},'' {\em
  Nucl.Phys.} {\bf B812} (2009) 1--11,
  \href{http://xxx.lanl.gov/abs/0808.0360}{{\tt 0808.0360}}.

\bibitem{Amariti:2009rb}
A.~Amariti, D.~Forcella, L.~Girardello, and A.~Mariotti, ``{3D Seiberg-like
  Dualities and M2 Branes},'' {\em JHEP} {\bf 1005} (2010) 025,
  \href{http://xxx.lanl.gov/abs/0903.3222}{{\tt 0903.3222}}.

\bibitem{Jafferis:2010un}
D.~L. Jafferis, ``{The Exact Superconformal $R$-Symmetry Extremizes $Z$},''
  \href{http://xxx.lanl.gov/abs/1012.3210}{{\tt 1012.3210}}.

\bibitem{Henningson:1998gx}
M.~Henningson and K.~Skenderis, ``{The holographic Weyl anomaly},'' {\em JHEP}
  {\bf 07} (1998) 023, \href{http://xxx.lanl.gov/abs/hep-th/9806087}{{\tt
  hep-th/9806087}}.

\bibitem{Klebanov:2002ja}
I.~R. Klebanov and A.~M. Polyakov, ``{AdS dual of the critical $O(N)$ vector
  model},'' {\em Phys. Lett.} {\bf B550} (2002) 213--219,
  \href{http://xxx.lanl.gov/abs/hep-th/0210114}{{\tt hep-th/0210114}}.

\bibitem{Aganagic:2002wv}
M.~Aganagic, A.~Klemm, M.~Marino, and C.~Vafa, ``{Matrix model as a mirror of
  Chern-Simons theory},'' {\em JHEP} {\bf 0402} (2004) 010,
  \href{http://xxx.lanl.gov/abs/hep-th/0211098}{{\tt hep-th/0211098}}.

\bibitem{Suyama:2009pd}
T.~Suyama, ``{On Large $N$ Solution of ABJM Theory},'' {\em Nucl.Phys.} {\bf
  B834} (2010) 50--76, \href{http://xxx.lanl.gov/abs/0912.1084}{{\tt
  0912.1084}}.

\bibitem{Drukker:2009hy}
N.~Drukker and D.~Trancanelli, ``{A Supermatrix model for $\mathcal N=6$ super
  Chern-Simons-matter theory},'' {\em JHEP} {\bf 1002} (2010) 058,
  \href{http://xxx.lanl.gov/abs/0912.3006}{{\tt 0912.3006}}.

\bibitem{Klebanov:1998hh}
I.~R. Klebanov and E.~Witten, ``{Superconformal field theory on threebranes at
  a Calabi-Yau singularity},'' {\em Nucl. Phys.} {\bf B536} (1998) 199--218,
  \href{http://xxx.lanl.gov/abs/hep-th/9807080}{{\tt hep-th/9807080}}.

\bibitem{Gubser:1998ia}
S.~Gubser, N.~Nekrasov, and S.~Shatashvili, ``{Generalized Conifolds and 4d
  $\mathcal N=1$ SCFT},'' {\em JHEP} {\bf 05} (1999) 003,
  \href{http://xxx.lanl.gov/abs/hep-th/9811230}{{\tt hep-th/9811230}}.

\bibitem{Benna:2008zy}
M.~Benna, I.~Klebanov, T.~Klose, and M.~Smedback, ``{Superconformal
  Chern-Simons Theories and $AdS_4$/CFT$_3$ Correspondence},'' {\em JHEP} {\bf
  0809} (2008) 072, \href{http://xxx.lanl.gov/abs/0806.1519}{{\tt 0806.1519}}.

\bibitem{Hosomichi:2008jd}
K.~Hosomichi, K.-M. Lee, S.~Lee, S.~Lee, and J.~Park, ``{${\cal N}=4$
  Superconformal Chern-Simons Theories with Hyper and Twisted Hyper
  Multiplets},'' {\em JHEP} {\bf 0807} (2008) 091,
  \href{http://xxx.lanl.gov/abs/0805.3662}{{\tt 0805.3662}}.

\bibitem{Franco:2009sp}
S.~Franco, I.~R. Klebanov, and D.~Rodriguez-Gomez, ``{M2-branes on Orbifolds of
  the Cone over $Q^{1,1,1}$},'' {\em JHEP} {\bf 08} (2009) 033,
  \href{http://xxx.lanl.gov/abs/0903.3231}{{\tt 0903.3231}}.

\bibitem{Davey:2009sr}
J.~Davey, A.~Hanany, N.~Mekareeya, and G.~Torri, ``{Phases of M2-brane
  Theories},'' {\em JHEP} {\bf 06} (2009) 025,
  \href{http://xxx.lanl.gov/abs/0903.3234}{{\tt 0903.3234}}.

\bibitem{Kontsevich:1992}
M.~Kontsevich, ``{Intersection theory on the moduli space of curves and the
  matrix Airy function},'' {\em Commun. Math. Phys.} {\bf 147} (1992) 1--23.

\bibitem{Santamaria:2010dm}
R.~C. Santamaria, M.~Marino, and P.~Putrov, ``{Unquenched flavor and tropical
  geometry in strongly coupled Chern-Simons-matter theories},''
  \href{http://xxx.lanl.gov/abs/1011.6281}{{\tt 1011.6281}}.

\bibitem{Bergman:2001qi}
A.~Bergman and C.~P. Herzog, ``{The volume of some non-spherical horizons and
  the AdS/CFT correspondence},'' {\em JHEP} {\bf 01} (2002) 030,
  \href{http://xxx.lanl.gov/abs/hep-th/0108020}{{\tt hep-th/0108020}}.

\end{thebibliography}\endgroup

\end{document}